\documentclass[prb,twocolumn,10pt,superscriptaddress,aps,amsmath,amsfonts,notitlepage,longbibliography]{revtex4-2}
\usepackage{amsmath,amsfonts,amssymb,dsfont,graphicx,bm}
\usepackage{color}
\usepackage{hyperref}
\usepackage{pbox}
\usepackage{balance}

\usepackage{txfonts}

\usepackage[normalem]{ulem}

\def\be{\begin{equation}}
\def\ee{\end{equation}}
\def\bea{\begin{eqnarray}}
\def\eea{\end{eqnarray}}
\def\bpm{\begin{pmatrix}}
\def\epm{\end{pmatrix}}

\def\Im{\mathop{\rm Im}}
\def\Re{\mathop{\rm Re}}

\begin{document}
\title{Dynamical response theory of interacting Majorana fermions \\ and its application to generic Kitaev quantum spin liquids in a field}

\author{Peng Rao}
\affiliation{Physics Department, Technical University of Munich, TUM School of Natural Sciences, 85748 Garching, Germany}
\author{Roderich Moessner}
\affiliation{Max Planck Institute for the Physics of Complex Systems, 01187 Dresden, Germany}
\author{Johannes Knolle}
\affiliation{Physics Department, Technical University of Munich, TUM School of Natural Sciences, 85748 Garching, Germany}
\affiliation{Munich Center for Quantum Science and Technology (MCQST), Schellingstr. 4, 80799 München, Germany}
\affiliation{Blackett Laboratory, Imperial College London, London SW7 2AZ, United Kingdom}

\date{\today}
\begin{abstract}
Motivated by the appearance of Majorana fermions in a broad range of correlated and topological electronic systems,
we develop a general method to compute the dynamical response of interacting Majorana fermions in the random-phase approximation (RPA).  This can be applied self-consistently on top of Majorana mean-field theory (MFT) backgrounds, thereby in particular providing a powerful tool to analyse {\it generic} behaviour in the vicinity of (various heavily studied) exactly soluble models. Prime examples are quantum spin liquids (QSL) with emergent Majorana excitations, with the celebrated exact solution of  Kitaev. 
We employ the RPA  to study in considerable detail phase structure and dynamics of the extended Kitaev honeycomb $KJ\Gamma$-model, with and without an applied field. 
First, we benchmark our method with Kitaev's exactly soluble model, finding a remarkable agreement. The interactions between Majorana fermions even turn out to mimic the effect of local $\mathbb{Z}_2$ flux excitations, which we explain analytically. Second, we  show how small non-Kitaev couplings $J$ and $\Gamma$  induce Majorana bound states, resulting in sharp features in the dynamical structure factor in the presence of fractionalisation: such 'spinon excitons' naturally appear, and can coexist and interact with the broad Majorana continuum. Third, for increasing couplings or field, our theory predicts instabilities of the KQSL triggered by the condensation of the sharp modes. From the high symmetry momenta of the condensation we can deduce which magnetically ordered phases surround the KQSL, in good agreement with previous finite-size numerics. We discuss implications for experiments and the broad range of applicability of our method to other QSL and Majorana systems. 
\end{abstract}

\maketitle

\section{Introduction}

Majorana particles are neutral fermionic excitations which are their own anti-particles. Although they were first proposed in particle physics, Majorana fermions have also become prominent in condensed matter physics, as they appear as low-energy excitations in a large number of theoretical systems such as the 2D Ising model~\cite{schultz1964ising,dotsenko1988fermion}, spinless $p$-wave superconductors in 1D (i.e. Kitaev wires)~\cite{kitaev2001unpaired} and 2D~\cite{ivanov2001non}, and fractional Quantum Hall states~\cite{read2000paired}. 
Furthermore, spin operators can be conveniently represented as parton Majorana fermions ~\cite{berezin1977particle,tsvelik1992new,shastry1997majorana,Fu2018,schaden2023bilinear}, which are particularly useful in studying strongly correlated spin systems, for example in the Kondo effect and heavy fermion Kondo lattices~\cite{vieira1981kondo,coleman1993kondo,tsvelik2007qft}.

In particular, parton Majorana fermions are instrumental in the study of $\mathbb{Z}_2$ quantum spin liquids (QSL)~\cite{tsvelik1992new,biswas20112,chen2012majorana},  exotic phases of quantum magnets where conventional magnetic order is destroyed by quantum fluctuations leading to long-range entangled ground states~\cite{balents_spin_2010,knolle2019field}. Here Majorana fermions arise naturally as fractionalized spin excitations in an emergent $\mathbb{Z}_2$ gauge field. However, for a long time progress in the field was hampered by the strongly correlated nature of QSL phases. Therefore, the advent of the Kitaev honeycomb model was a breakthrough development~\cite{Kitaev2006}, as it provides an exactly soluble example of Majorana excitations in a gapped $\mathbb{Z}_2$ gauge field emerging from interacting spins-$1/2$. 

One of the most promising developments towards experimental detection of Majorana fermions was the proposal that the Kitaev model's bond-dependent Ising interactions on the honeycomb lattice can be realized in spin-orbit coupled Mott insulators~\cite{Jackeli2009}, e.g. the iridate compounds A$_2$IrO$_3$ (A$=$ Na, Li) and in particular $\alpha$-RuCl$_3$~\cite{plumb2014alpha}. Over the last decade, there has been intensive search for Majorana QSL signatures in such Kitaev materials, for reviews see Refs.\cite{hermanns2018physics,takagi2019concept,motome2020hunting,Trebst2022,rousochatzakis2024beyond,matsuda2025kitaevquantumspinliquids}. Another important direction is the theoretical determination of Majorana signatures in experimental probes, for example, dynamical response functions as measured in inelastic scattering experiments which can only be calculated exactly for the pure Kitaev model~\cite{Knolle2014,knolle2015dynamics,knolle2014raman,knolle2016dynamics}. 

\begin{figure*}
    \centering
    \includegraphics[width=\linewidth]{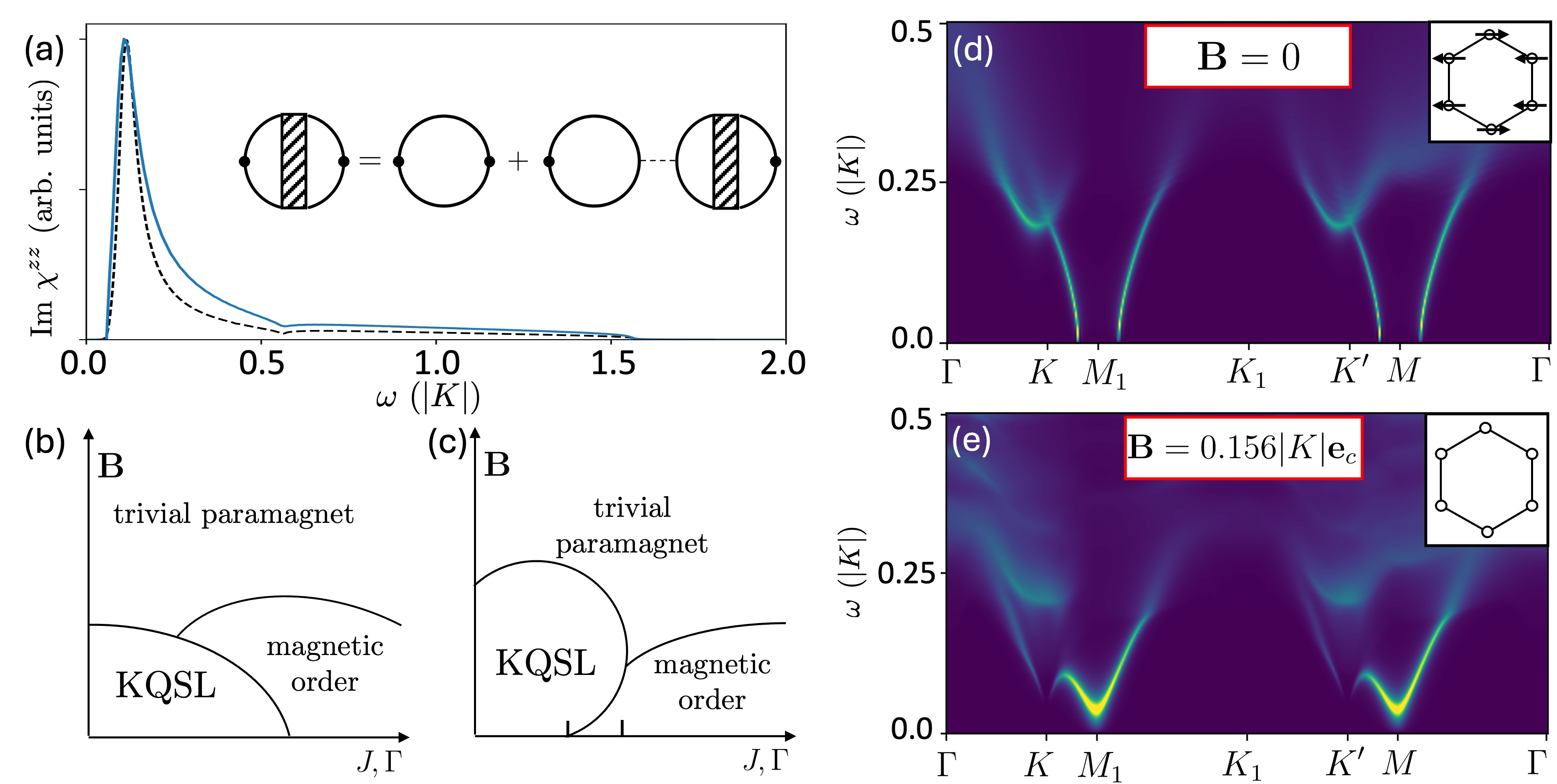}
    \caption{(a) The self-consistent Majorana RPA susceptibility $\Im \chi^{zz}$ (blue line) is shown as a function of frequency at $\mathbf{k}=0$ for the ferromagnetic Kitaev model and compared to the exact response (dotted line with intensity normalized to have the same peak intensity). For comparison, $\Im \chi^{zz}$ is truncated at the peak threshold to remove finite broadening effects; the threshold is shifted to the exact vison gap $\Delta_v\approx 0.07 |K|$, see discussion in Sec.~\ref{sec:results-Kitaev}. The inset: diagrammatic representation of the RPA approximation for the spin susceptibility $\chi^{\alpha\beta}(\omega,\mathbf{k})$. (b)-(c): Schematic phase diagrams at finite field. In (b) the KQSL is stable at small $J,\Gamma$ and $\mathbf{B}$, which all destabilize the KQSL for larger values. In (c) within certain $J,\Gamma$ intervals, the system is magnetically ordered at zero field but the KQSL is stabilized at finite field. (d)-(e) INS intensity for the field-induced KQSL at FM $K$, $J=0.11|K|, \Gamma = 0$ and: (d) $\mathbf{B} = 0$ where the system is in the stripy phase (inset); (e) perpendicular $\mathbf{B} = 0.156|K|\mathbf{e}_c$. The $M$-, $M_1$-modes that condense at zero field are lifted to finite fields and broaden, suggesting a field-induced KQSL. For the Brillouin zone (BZ) conventions, see Fig.~\ref{fig:lattice}(c).}
    \label{fig:schematic}
\end{figure*}

Unfortunately, determining the dynamical response of Majorana fermions in a given setting is hampered by the inevitable presence of interactions beyond ideal free Majorana Hamiltonians. In our example of Kitaev systems, beyond the pure Kitaev point additional exchange interactions spoil the exact solubility, leading to an interactiong Majorana fermion problem. Indeed, all candidate materials display long-range magnetic order because additional spin interactions beyond Kitaev assert themselves. However, in the candidate material $\alpha$-RuCl$_3$, the application of an intermediate in-plane magnetic field suppresses the observed zig-zag magnetism~\cite{kubota2015successive,wolter2017field,sears2017phase} and it was argued that a distinct intermediate field phase in the range $\approx 8 - 12$ Tesla appears before the high-field polarized ferromagnet. There, unconventional excitations have been interpreted in terms of Majorana excitations of a putative chiral Kitaev QSL (KQSL)~\cite{banerjee2018excitations,wang2017magnetic,li2021identification,ponomaryov2020nature,wulferding2020magnon,wang2020range,sahasrabudhe2024chiral}. Moreover, signatures of a thermal Hall conductivity reminiscent of the one predicted for the soluble Kitaev model in a field~\cite{kasahara2018majorana,bruin2022robustness,chern2021sign} point to an unconventional phase, but the nature of the excitations is under debate~\cite{winter2017breakdown,czajka2023planar}.

A related challenge is the absence of an established microscopic spin Hamiltonian for Kitaev materials~\cite{Maksimov2020}. More generally, this necessitates calculating dynamical response for \textit{generic} interacting Majorana systems beyond the soluble point. In the case of KQSL, on the one hand, numerical approaches like exact diagonalization (ED)~\cite{winter2017breakdown} and DMRG~\cite{Gohlke2017,Gohlke2018} have been employed but suffer from strong finite size effects. On the other hand, perturbative approaches trying to incorporate the dynamics of flux excitations from additional interactions have been developed but are either somewhat ad-hoc, or only valid close to the soluble point~\cite{tikhonov2011power,Song2016,Knolle2018,Zhang2021,Joy2022,chen2023nature,cookmeyer2023dynamics}.

This paper pursues a pair of complementary objectives. First, we develop a general method for computing dynamical response functions of a general interacting Majorana Hamiltonian in the random phase approximation (RPA). Our method can be applied \textit{self-consistently} to parton mean field theories (MFT) which are often employed in studying QSLs~\cite{anderson1987resonating,anderson1987resonating-1}. Here the RPA includes to leading order interactions between Majoranas on top of the MF background. 

Second, we use the formalism thus developed to study the dynamical spin structure factor of the KQSL for generic extended Kitaev models in finite magnetic fields, thereby providing predictions for inelastic neutron scattering (INS) experiments. In the absence of a microscopic Hamiltonian and controlled analytical methods, our formalism addresses the following general questions: How can additional spin interactions beyond Kitaev qualitatively alter the dynamical response of the KQSL in a field? How to incorporate interaction effects beyond the parton mean-field description for calculating dynamical response functions of Majorana fermions?

Taken together, these two strands extend the reach of Majorana descriptions of correlated quantum matter in general, and in particular evince a rich phenomenology of generically perturbed Kitaev spin liquids in a unified framework. 

\subsection{Overview}
We first present a very general form of the dynamical response theory starting with a susceptibility for an arbitrary Majorana bilinear vertex (e.g. for charge or spin) under a generic interacting Majorana Hamiltonian. This is accomplished by the RPA formalism which sums over the geometric series of one-loop Majorana fermion diagrams; see Fig.~\ref{fig:schematic}(a) inset. Although Majorana fermions can be equivalently written as linear combinations of complex fermions, evaluating dynamical response functions in the Majorana basis is particularly convenient for systems in which Majorana fermions appear naturally as elementary excitations. As a key application, we apply our method to the KQSL in the $KJ\Gamma$-model which, apart from the Kitaev Hamiltonian, also includes nearest neighbour (N.N.) Heisenberg and off-diagonal spin couplings $J$ and $\Gamma$. The starting point for our calculations is a parton MFT decoupling of the Hamiltonian leading to a Majorana bandstructure with the $\mathbb{Z}_2$ flux excitations only appearing indirectly as additional parton bands~\cite{Knolle2018}. 

We then go beyond MFT and compute the spin susceptibility by considering the interactions of Majorana fermions on top of the MFT background. First, we benchmark our self-consistent Majorana RPA with the exact solution of the pure Kitaev limit and find, somewhat surprisingly, that it reproduces almost quantitatively the exact spin susceptibility~\cite{Knolle2014}, see Fig.~\ref{fig:schematic}(a) and Sec.~\ref{sec:results-Kitaev}. Given that flux excitations, which are absent in our MFT description, are crucial in the exact solution for obtaining the correct structure factor, the agreement is remarkable. From the analytical structure of the RPA at the soluble point, we can understand this by showing how interactions between Majorana fermions mimic the effect of local flux excitations. In this context we note that the self-consistent mean-field+RPA theory is a well established method for describing itinerant anti-ferromagnets~\cite{Schrieffer1989,chubukov1992renormalized,knolle2010theory,knolle2011multiorbital,willsher2023magnetic} and in Ref.~\cite{Ho2001} \textit{Ho, Muthukumar, Ogata and Anderson} employed it for the first time to study the effect of spinon excitations on top of a parton MFT. Here we establish that it can be a powerful yet simple method for calculating dynamical response functions {\it within} QSL phases. Indeed, a similar approach for the Dirac QSL in the $J_1-J_2$ model on the triangular lattice~\cite{willsher2025dynamics} is in good agreement with recent DMRG~\cite{drescher2023dynamical} and variational Monte Carlo results~\cite{ferrari2019dynamical}.  

Away from the Kitaev limit at small $J, \Gamma$ and zero field, we find that the broad continua of the spin susceptibility change significantly and may display spin-flip excitations sharp in energy. Such `para-magnon'  modes are collective excitations of Majorana fermions and within the RPA scheme appear as distinct excitonic bound states. The identification of these modes has two major implications. First, they represent qualitatively new excitations in addition to the two-Majorana continuum. Second, their condensation can also be used to infer the phase diagram of the extended $KJ\Gamma$-model. For example, the modes are centered around specific regions in the Brillouin zone (BZ) and their condensation at particular momenta indicate instabilities of the KQSL towards magnetic long-range ordered phases. 
Remarkably, the phase diagram from mode condensation is in qualitative agreement with numerical exact diagonalization (ED)~\cite{Rau2014}, and DMRG results~\cite{Gohlke2017,Gohlke2018}.

Qualitatively, finite magnetic fields generally tend to weaken the KQSL and cause one of the existing para-magnon modes to condense, as shown schematically in Fig.~\ref{fig:schematic}(b). However, we also find the opposite regime where the magnetic field {\it stabilizes} the KQSL for specific parameter regimes: the sharp modes that are condensed at zero magnetic field can be gapped at intermediate field values resulting in a stable KQSL phase; see Figs.~\ref{fig:schematic}(c)-(e). Thus, we find within our method an explicit illustration of existence and mechanism of a `field-induced KQSL'. Fig.~\ref{fig:phase-schematic} summarizes results at both zero and finite field.

The rest of the paper is organized as follows. In Sec.~\ref{sec:RPA} we develop the general theory for computing RPA dynamical response of an interacting Majorana system. In Sec.~\ref{sec:model}, we introduce the $KJ\Gamma$-model and the MFT approximation. We then adapt the general RPA formalism to compute specifically the dynamic spin susceptibility. In Sec.~\ref{sec:results-Kitaev} we consider the pure Kitaev model at zero magnetic field and compare the self-consistent RPA results with the spin structure factor of the exact solution. The results for the $KJ\Gamma$-model at zero field are presented in Sec.~\ref{sec:results-KJGamma} including the transitions to competing ordered phases. Finally in Sec.~\ref{sec:results-finite-field}, we discuss the effect of finite external fields on the KQSL. In particular, the `field-induced KQSL' scenario is demonstrated explicitly. Sec.~\ref{sec:concl} closes with a concluding discussion.

\section{\label{sec:RPA} Dynamic susceptibilities of general interacting Majorana fermions}

In this section we develop the self-consistent RPA theory of the dynamical response for interacting Majorana systems. To maximise the utility of our treatment, also beyond the case of Kitaev QSLs covered extensively below, we develop the formalism for arbitrary Majorana fermions $\gamma_a(\mathbf{r})$, where $a$ are internal state indices; generally we have $a=1,...,2N$ where $N$ is a positive integer. In our specific example of the KQSL, they describes the collection of Majorana fermions $\gamma=(c,b^x,b^y,b^z)$ on each site; see Eq.~\eqref{eq:Majorana-rep} below. 

\subsection{Model}

\begin{figure}
    \centering
    \includegraphics[width=\linewidth]{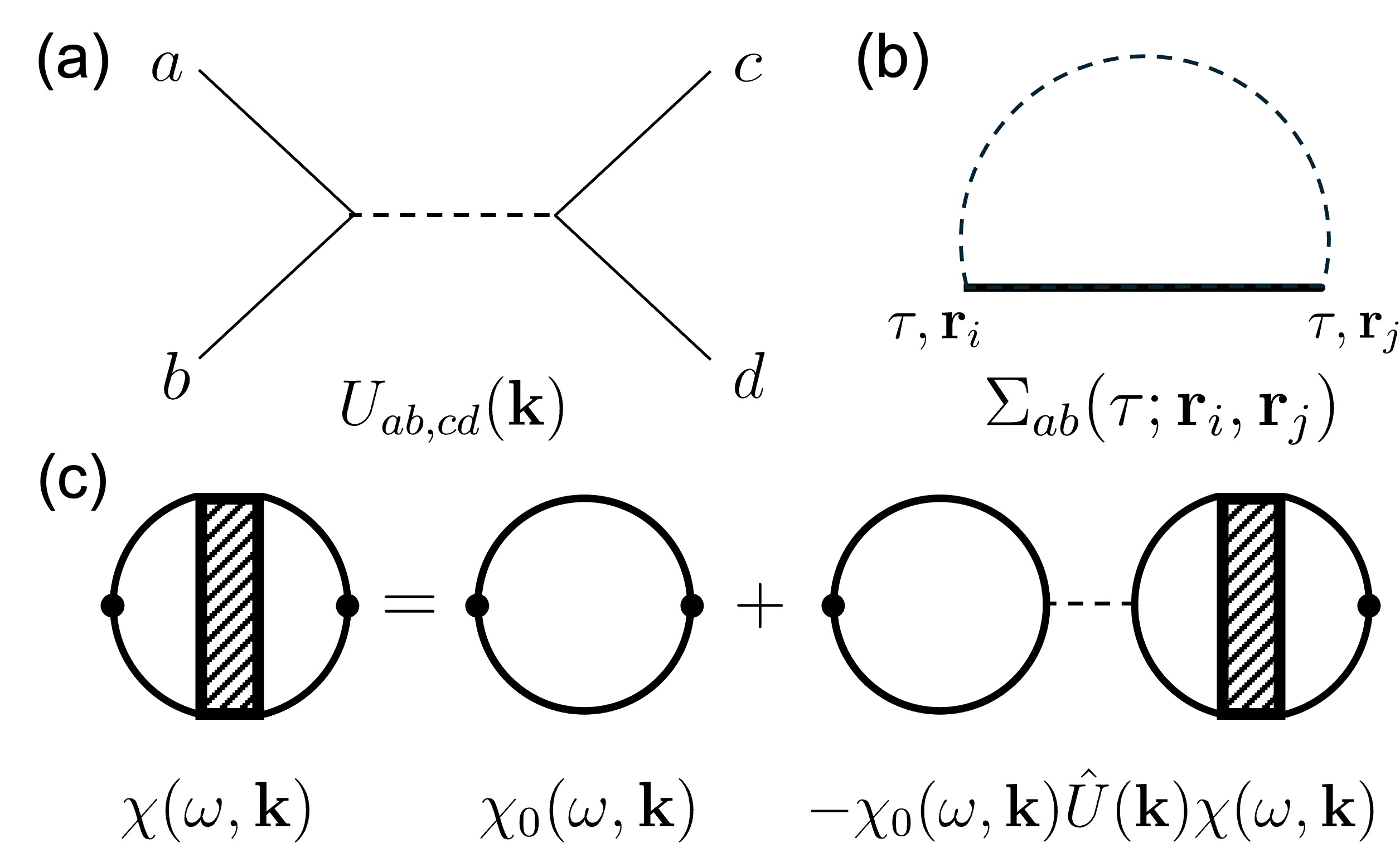}
    \caption{Diagrammatic representation of the RPA spin susceptibility. (a) the four-Majorana interaction vertex given by Eq.~\eqref{eq:generalised-interaction-vertex}. (b) the Majorana self-energy in real space in the self-consistent Born approximation (SCBA). The thick line denotes the renormalized Green's function. (c) The RPA approximation for the generalized susceptibility in the Majorana basis given by Eq.~\eqref{eq:generalised-one-loop}. The `symmetrized' interaction vertex $\hat{U}(\mathbf{k})$ is given by Eq.~\eqref{eq:interactoin-vertex-symmetrised}.
    }
    \label{fig:susceptibility-diagram}
\end{figure}

We study a general interacting Majorana Hamiltonian
\begin{equation}
    H=H_0+V,
\end{equation}
which includes the single-particle Hamiltonian
\begin{equation}
    H_0 =\frac{i}{2} \sum_{a,b} \sum_{\mathbf{p}} H_{ab}(\mathbf{p}) \overline{\gamma}_{a}(\mathbf{p}) \gamma_{b}(\mathbf{p}),
    \label{eq:free-Hamiltonian-momentum}
\end{equation}
and Majorana interactions $V$ which can be written in real space as
\begin{equation}
    V =- \sum_{i,j} \sum_{a\ne b, c\ne d} U_{ab,cd}(\mathbf{r}_i-\mathbf{r}_j) \gamma_a(\mathbf{r}_i) \gamma_c(\mathbf{r}_j) \gamma_d(\mathbf{r}_j) \gamma_b(\mathbf{r}_i).
\end{equation}
In momentum space this gives
\begin{equation}
    V = -\sum_{\mathbf{p},\mathbf{p}',\mathbf{k}}U_{ab,cd}(\mathbf{k})\overline{\gamma}_{a}(\mathbf{p}-\mathbf{k})\overline{\gamma}_{c}(\mathbf{p}'+\mathbf{k})\gamma_{d}(\mathbf{p}')\gamma_{b}(\mathbf{p}),
    \label{eq:generalised-interaction-vertex}
\end{equation}
where $U_{ab,cd}(\mathbf{k})$ are matrix elements of the interaction vertices shown in Fig.~\ref{fig:susceptibility-diagram}(a). Note that for interactions between Majorana fermions on different sites, the issue of normal ordering does not arise. 

We choose the following convention for Majorana anti-commutation relations
\begin{equation}
    \{\gamma_a(\bm{r}_i),\gamma_b(\mathbf{r}_j)\}=\delta_{ab}\delta_{ij}
\end{equation}
or in momentum space
\begin{equation}
\{\gamma_a(\mathbf{p}) , \overline{\gamma}_b(\mathbf{p}')  \} = (2\pi)^2\delta_{ab} \delta(\mathbf{p}-\mathbf{p}').
\label{eq:Majorana-definition}
\end{equation}
Here $\overline{\gamma}_{a}(\mathbf{p})  = [\gamma_{a}(\mathbf{p})]^\dagger$ and we have $(2\pi)^2\sum_\mathbf{p} \delta(\mathbf{p})=1$. The Majorana operators satisfy the identity
\begin{equation}
      \overline{\gamma}_{a}(\mathbf{p}) =  \gamma_{a}(-\mathbf{p}).\label{eq:Majorana-identity}
\end{equation}
It follows from hermiticity of $H_0$ and Eq.~\eqref{eq:Majorana-identity} that 
\begin{equation}
    H_{ab}^*(\mathbf{p}) = - H_{ba}(\mathbf{p}) = H_{ab}(-\mathbf{p}).\label{eq:Hamiltonian-identity}
\end{equation}

As a first step, we find the non-interacting Majorana bandstructure by diagonalizing \eqref{eq:free-Hamiltonian-momentum}. This can be done using the Bogoliubov transformation of $\gamma_a(\mathbf{p})$ into complex fermions $\alpha_n(\mathbf{p})$ with energy level indices $n=1,..,N$. Due to Eq.~\eqref{eq:Majorana-identity} the transformation has the form:
\begin{equation}
    \gamma_{a}(\mathbf{p}) =u_{\alpha n }(\mathbf{p}) \alpha_n(\mathbf{p})  + u_{\alpha n}^*(-\mathbf{p}) \alpha_n^\dagger(-\mathbf{p}) ,\label{eq:Bogoliubov-transform}
\end{equation}
The coefficients $u_{\alpha n }(\mathbf{p})$ are normalized to be $\sum_n[|u_{\alpha n }(\mathbf{p})|^2 + |u_{\alpha n }(-\mathbf{p})|^2]=1$ such that the complex fermions satisfy the standard anti-commutation relation:
\begin{equation}
\begin{split}
\{\alpha_n(\mathbf{p}) , \alpha_m^\dagger(\mathbf{p}')  \} &= (2\pi)^2\delta_{mn} \delta(\mathbf{p}-\mathbf{p}').
\end{split}
\end{equation}
The identity follows from the mapping Eq.~\eqref{eq:Majorana-definition}. The coefficients $u_{\alpha n }(\mathbf{p})$ can be found from the requirement that the free Hamiltonian $H_0$ has the form:
\begin{equation}
    H_{0} = \sum_{n=1}^N \sum_{\mathbf{p}}E_n(\mathbf{p}) \left[\alpha_n^\dagger(\mathbf{p})\alpha_n(\mathbf{p}) -\frac{1}{2} \right]; \  \  E_n(\mathbf{p}) >0.\label{eq:Hamiltonian-bogoliubov}
\end{equation}
By substituting Eqs.~\eqref{eq:free-Hamiltonian-momentum} and \eqref{eq:Bogoliubov-transform} into the equation of motion $[H_{0} , \alpha_n^\dagger(\mathbf{p})] = E_n(\mathbf{p})\alpha_n^\dagger(\mathbf{p})$, we obtain the eigenvalue problem:
\begin{eqnarray}
    iH_{ab}(\mathbf{p}) u_{b n}(\mathbf{p}) &=& E_n(\mathbf{p}) u_{a n} (\mathbf{p}). \\
    iH_{ab}(\mathbf{p}) u_{b n}^*(-\mathbf{p}) &=&-E_n(-\mathbf{p}) u_{a n}^* (-\mathbf{p}).
\end{eqnarray}
Therefore, for a particle state with energy $E_n(\mathbf{p})>0$ and eigenstate $u_{a n} (\mathbf{p})$, we have the corresponding hole eigenstate $u_{a n}^* (-\mathbf{p})$ with energy $-E_n(-\mathbf{p})<0$. This can be also directly verified by substituting Eq.~\eqref{eq:Bogoliubov-transform} into Eq.~\eqref{eq:free-Hamiltonian-momentum}, which gives Eq.~\eqref{eq:Hamiltonian-bogoliubov} as it should. In this paper we shall use $n>0$ to denote the particle branch of the spectrum, and only take its matrix elements $u_{an}$ to avoid phase ambiguity.

\subsection{Majorana Green's functions}

The Majorana Green's functions are defined as:
\begin{equation}
G_{ab}(\tau,\mathbf{p}) =- \langle \text{T}_\tau \{ \gamma_a(\tau,\mathbf{p}) \overline{\gamma}_b(0,\mathbf{p})\} \rangle.\label{eq:Greens-function-def}
\end{equation}
Here $\text{T}_\tau$ is the time-ordering with respect to imaginary time $\tau$.
The 'complex conjugate' Majorana fermion $\overline{\gamma}_a(\tau,\mathbf{p})$ is
\begin{equation}
  \overline{\gamma}_a(\tau,\mathbf{p}) = e^{\tau H}\overline{\gamma}_a(\mathbf{p}) e^{-\tau H}.
\end{equation}
Eq.~\eqref{eq:Greens-function-def} can be expressed in terms of coefficients $u_{an}$ after going over to the Matsubara frequency, by substituting \eqref{eq:Bogoliubov-transform} then performing the average using the Hamiltonian \eqref{eq:Hamiltonian-bogoliubov}: 
\begin{equation}
G_{ab} (i\omega_l ,\mathbf{p}) =
\sum_n\left[\frac{u_{a n }(\mathbf{p})u_{b n }^*(\mathbf{p})}{i\omega_l -E_n(\mathbf{p})}  + \frac{u_{a n}^*(-\mathbf{p}) u_{b n}(-\mathbf{p})}{i\omega_l +E_n(-\mathbf{p})} \right].\label{eq:Greens-functions}
\end{equation}
The fermionic Matsubara frequency is $\omega_l = (2l+1) \pi T$. Note that the Matsubara Green's functions satisfy
\begin{subequations}
\begin{align}
[G_{ab}(i\omega_l,\mathbf{p})]^* &=-G_{ab}(i\omega_l,-\mathbf{p}), \\
G_{ab}(i\omega_l,\mathbf{p}) &= -  G_{ba}(-i\omega_l,-\mathbf{p}), 
\end{align}\label{eq:Greens-function-identity}
\end{subequations}
which follows from
\begin{equation}\label{eq:Majorana-identity-Matsubara}
\gamma_a(\tau,\mathbf{p}) = \overline{\gamma}_a(\tau,-\mathbf{p}), \ [\gamma_a(\tau,\mathbf{p})]^\dagger = \overline{\gamma}_a(-\tau,\mathbf{p}).
\end{equation}

For Majorana MFT, e.g. commonly used in parton descriptions of QSL like the KQSL, the single-particle $H_0$ itself is obtained from decouping the interactions $V$. It can be shown directly that the MFT Majorana Green's functions corresponds to the self-consistent Born approximation (SCBA) for the Majorana self-energies, which sums all diagrams in which the interaction lines do not cross; see Fig.~\ref{fig:susceptibility-diagram}(b) and Appendix~\ref{sec:MajoranaMF:SCBA} for an example of the KQSL. Thus our method can be applied self-consistently to Majorana MFT.

\subsection{Generalized susceptibility}

We  consider the dynamical response of Majorana fermions coupled to external perturbations via vertices of the form
\begin{equation}
    O_i(\mathbf{r}) = i \sum_{a,b}\Gamma_{ab}^{(i)} \gamma_a (\mathbf{r})\gamma_b (\mathbf{r}).
    \label{eq:generalised-vertex}
\end{equation}
For example, in the KQSL the spin operator has this form, see Eq.~\eqref{eq:Majorana-rep} below. The generalized susceptibility can be defined in the Matsubara representation as:
\begin{equation}
    \chi_{ij}(\omega,\mathbf{k}) = \lim_{i\omega_0 \to \omega +i\delta }\sum_{k,l} \int \mathrm{d}\tau P_{ij}(\tau,\mathbf{r}_{k} - \mathbf{r}_{l}) e^{i \omega_0 \tau -i\mathbf{k}.(\mathbf{r}_{k} -\mathbf{r}_l)},\label{eq:generalised-susceptibility-Matsubara}
\end{equation}
with the time-ordered correlation function:
\begin{equation}
    P_{ij}(\tau,\mathbf{r}_{k}-\mathbf{r}_l) = \langle \text{T}_\tau \{O_i(\tau,\mathbf{r}_k)O_j(0,\mathbf{r}_l) \}\rangle.
\end{equation}

We compute Eq.~\eqref{eq:generalised-susceptibility-Matsubara} in the RPA approximation at zero temperature $T=0$, which consists of summing over one-loop Majorana diagrams. To obtain closed expressions for the RPA for a general momentum-dependent interaction vertex $U_{ab,cd}(\mathbf{k})$,  we sum only over diagrams shown in Fig.~\ref{fig:susceptibility-diagram}(c) with constant four-momentum transfer $k=(i\omega_0,\mathbf{k})$ along the interaction lines. In these diagrams the internal loop momenta do not `spill into' each other so that the result factorizes into a geometric series of one-loop diagrams. However, we note that if the interaction is on-site, i.e. momentum-independent, then ladder-type diagrams also factorize and contribute to the RPA. We do not consider these here but their inclusion is straightforward.

The one-loop tensor as shown in Fig.~\ref{fig:susceptibility-diagram}(c) is:
\begin{widetext}
\begin{subequations} \label{eq:generalised-susceptibility-one-loop}
\begin{align}
    &[\chi_0(\omega,\mathbf{k})]_{ab,cd}  = \lim_{i\omega_0 \to \omega +i\delta }\sum_{k,l} \int \mathrm{d}\tau P^{(0)}_{ab,cd}(\tau,\mathbf{r}_{k} -\mathbf{r}_l) e^{i \omega_0 \tau -i\mathbf{k}.(\mathbf{r}_{k} -\mathbf{r}_l)}; \\
    &P^{(0)}_{ab,cd}(\tau,\mathbf{r}) =-\langle \text{T}_\tau \{\gamma_a(\tau,\mathbf{r})\gamma_b(\tau,\mathbf{r}) \gamma_c(0,0) \gamma_d(0,0) \}\rangle_0,
\end{align}    
\end{subequations}
with averages taken over the non-interacting ground state after MFT decoupling. Eq.~\eqref{eq:generalised-susceptibility-one-loop} can be computed numerically taking the Bogoliubov coefficients $u_{an}$ in Eq.~\eqref{eq:Bogoliubov-transform} as input:
\begin{equation}\label{eq:generalised-one-loop}
\begin{split}
    [\chi_0(k)]_{ab,cd}=-T\sum_{\omega_l,\mathbf{p}}\bigg[G_{ac}(p+k)G_{db}(p)-G_{ad}(p+k)G_{cb}(p)\bigg].       
\end{split}   
\end{equation}
Here, we use the shorthand notation $k=(\omega_0,\mathbf{k})$ and have substituted Eqs.~\eqref{eq:Greens-function-identity} and \eqref{eq:Majorana-identity-Matsubara}. We now integrate over $\tau$ and substitute Eqs.~\eqref{eq:Greens-function-def} and \eqref{eq:Greens-functions}. The result can be expressed as a sum of terms, each of which containing a product of frequency poles. We then sum over the internal Matsubara frequency $\omega_l$ and take the limit $T\to 0$ (remember $E_n>0$). Finally, performing the analytical continuation $i\omega_0 \rightarrow \omega + i\delta$ gives
\begin{equation}
\begin{split}\label{eq:generalised-one-loop-result}
    &[\chi_0(k)]_{ab,cd}   
    =-\sum_{\mathbf{p}} \sum_{m,n} \Biggl[ \frac{u_{an}(\mathbf{p}+\mathbf{k})u_{cn}^{*}(\mathbf{p}+\mathbf{k})u_{dm}^{*}(-\mathbf{p})u_{bm}(-\mathbf{p})}{\omega -E_n(\mathbf{p}+\mathbf{k})-E_m(-\mathbf{p})+  i\delta}  
    -  \frac{u^{*}_{an}(-\mathbf{p}-\mathbf{k}) u_{cn}(-\mathbf{p}-\mathbf{k})u_{dm}(\mathbf{p})u^{*}_{bm}(\mathbf{p})}{\omega +E_n(-\mathbf{p}-\mathbf{k})+E_m(\mathbf{p})+  i\delta} \\
    &  - \frac{u_{an}(\mathbf{p}+\mathbf{k})u^{ *}_{d n}(\mathbf{p}+\mathbf{k})u^{*}_{cm}(-\mathbf{p})u_{bm}(-\mathbf{p})}{\omega -E_n(\mathbf{p}+\mathbf{k})-E_m(-\mathbf{p})+  i\delta} 
    + \frac{u^{*}_{an}(-\mathbf{p}-\mathbf{k}) u_{dn}(-\mathbf{p}-\mathbf{k})u_{cm}(\mathbf{p})u^{*}_{bm}(\mathbf{p})}{\omega +E_n(-\mathbf{p}-\mathbf{k})+ E_m(\mathbf{p})+ i\delta} \Biggl].
\end{split}
\end{equation}
\end{widetext}
The frequency poles at $\omega \pm (E_n+E_m)$ represent the emission and absorption of Majorana pairs. 
   
For the RPA susceptibility, the summation over the one-loop diagrams, Fig.~\ref{fig:susceptibility-diagram}(c) gives the  `matrix equation'
\begin{equation}
     [\chi(\omega,\mathbf{k})]_{ab,cd} = [\chi_0(\omega,\mathbf{k})]_{ab,kl}[1 + \hat{U} (\mathbf{k})\chi_0(\omega,\mathbf{k})]^{-1}_{kl,cd},\label{eq:generalised-RPA-tensor}
\end{equation}
where the interaction vertices are symmetrized with respect to the `row' and `column' indices:
\begin{equation}
    [\hat{U}(\mathbf{k})]_{ab,cd}  = U_{ab,cd} (-\mathbf{k}) + U_{cd,ab}(\mathbf{k}).\label{eq:interactoin-vertex-symmetrised}
\end{equation}
From the tensor \eqref{eq:generalised-RPA-tensor} one obtains 
\begin{equation}
    \chi_{ij}(\omega,\mathbf{k}) = \Gamma^{(i)}_{ab} [\chi(\omega,\mathbf{k})]_{ab,cd}\Gamma^{(j)}_{cd}.
    \label{eq:RPA-generalised-susceptibility}
\end{equation}
Eq.~\eqref{eq:RPA-generalised-susceptibility} then gives the RPA susceptibility of a general Majorana Hamiltonian with respect to the vertices in Eq.~\eqref{eq:generalised-vertex}.

\section{\label{sec:model}Extended Kitaev $KJ\Gamma$-model}

\begin{figure}
    \centering
    \includegraphics[width=\linewidth]{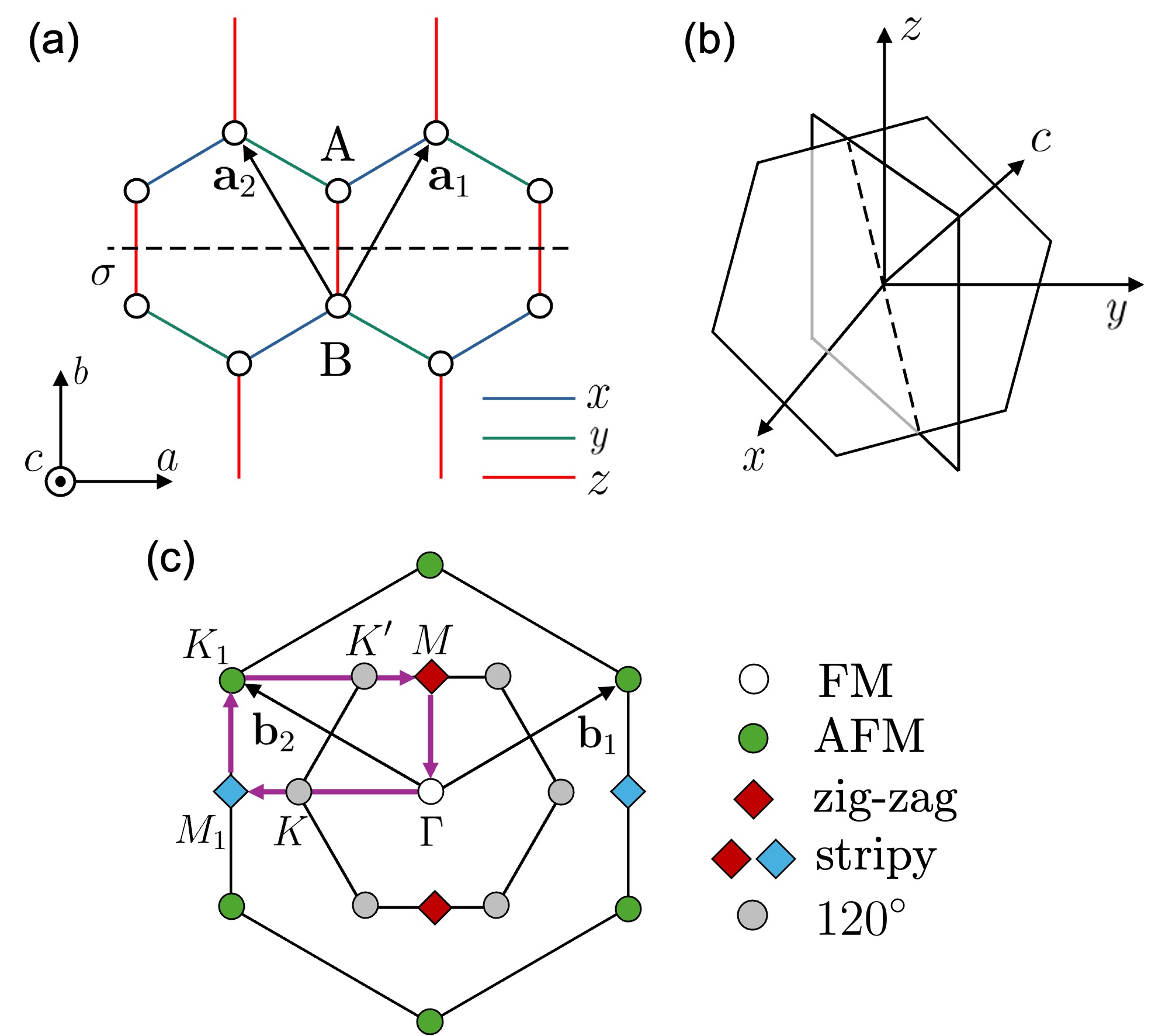}
    \caption{(a) The 2D honeycomb lattice of magnetic atoms on sublattices A and B. The $x$-, $y$-, $z$-bonds are shown in color. The lattice vectors $\mathbf{a}_1, \mathbf{a}_2$ and the local frame extended by the crystal axes are shown explicitly. The reflection plane $\sigma$ perpendicular to the $z$-bonds is shown as dashed lines. (b) The 2D honeycomb lattice embedded in the 3D lattice with global $x$-, $y$-, $z$-axes. (c) The BZ convention, where the first and second BZs are shown, as well as the high-symmetry points and Bragg peak positions of the magnetic orders. The reciprocal lattice vectors are $\mathbf{b}_1, \mathbf{b}_2$. The momentum path for the INS plots in this paper is shown in magenta.}
    \label{fig:lattice}
\end{figure}

In the rest of this paper, we apply our general formalism  to the $\mathbb{Z}_2$ Kitaev QSL in the $KJ\Gamma$-model. We first discuss the pure Kitaev model,  defined on a two-dimensional (2D) honeycomb lattice with N.N. anisotropic spin interactions:
\begin{equation}\label{eq:Kitaev-model}
H_K =  K \sum_{i,j \in \alpha} S^\alpha_i S^\alpha_j,
\end{equation}
where $S^\alpha$ are spin-$1/2$ operators, and the summation is taken over N.N. sites connected by $\alpha$-bonds as shown in Fig.~\ref{fig:lattice}(a). Later we focus on the $KJ\Gamma$-model with an external magnetic field
\begin{equation}\label{eq:spin-Hamiltonian}
    H = H_K + J\sum_{i,j}\mathbf{S}_i.\mathbf{S}_j+   \Gamma \sum_{\substack{i,j \in \alpha \\ \alpha \ne \beta \ne \gamma}} 
    \left(S^\beta_i S^\gamma_j +S^\gamma_i S^\beta_j\right) - \sum_i \mathbf{B}.\mathbf{S}_i,
\end{equation}
for both ferro-(FM) and antiferromagnetic(AFM) Kitaev couplings, and $J$ and $\Gamma$ are the N.N. Heisenberg and off-diagonal couplings respectively.

The symmetry properties of the spin Hamiltonian \eqref{eq:spin-Hamiltonian} are as follows. In candidate materials such as $\alpha$-RuCl$_3$, spin vectors are defined in the global laboratory frame whereas the honeycomb lattice lies on the $(111)$-surface; see Fig.~\ref{fig:lattice}(b). In the absence of magnetic field, the Hamiltonian \eqref{eq:spin-Hamiltonian} has the following symmetries: C$_6$-rotation around the center of the honeycomb followed by a lattice plane reflection; mirror reflection $\sigma$ across the bond; and time-reversal $T$. An external magnetic field generally breaks all symmetries of the system. However, for certain field directions, the system retains a nontrivial symmetry. For field perpendicular to the surface $\mathbf{B}\parallel(111)$, the system is still C$_6$ invariant. And for field along one of the bonds, e.g. the $z$-bond $\mathbf{B}\parallel(\bar{1}10)$, the system is invariant under reflection $\sigma$ with the bond being normal to the reflection plane. In this paper, we shall consider both field directions $\mathbf{B}\parallel(111)$ and $\mathbf{B}\parallel(\bar{1}10)$.

For reference, the local frame extended by the crystal axes is related to the global frame by:
\begin{equation}\label{eq:reference-frame}
\begin{pmatrix}
    \mathbf{e}_a \\ \mathbf{e}_b \\ \mathbf{e}_c      
\end{pmatrix}
= U
\begin{pmatrix}
    \mathbf{e}_x \\ \mathbf{e}_y \\ \mathbf{e}_z
\end{pmatrix}; \ 
U = \frac{1}{\sqrt{6}}\begin{pmatrix}
    1 & 1& -2 \\
    -\sqrt{3} & \sqrt{3} & 0 \\
    \sqrt{2}&\sqrt{2}&\sqrt{2}
\end{pmatrix}.
\end{equation}
Here $\mathbf{e}_{a}$ is the unit vector along the $a$-axis etc. and the 2D honeycomb lattice is extended by $\mathbf{e}_a$ and $\mathbf{e}_b$; see Fig.~\ref{fig:lattice}(a)-(b). In this paper, we shall use Greek letters to refer to vector components in the global frame. The high symmetry field directions are $\mathbf{e}_c$ for $(111)$ and $\mathbf{e}_b$ for $(\bar{1}10)$ respectively.

To study the KQSL in the spin Hamiltonian~\eqref{eq:spin-Hamiltonian}, the most physically transparent approach utilizes the Majorana representation of spin operators. Following Ref.~\cite{Kitaev2006}, we represent the spin operators at a given site $i$ in terms of four Majorana fermions
\begin{equation}\label{eq:Majorana-rep}
    S^\alpha_i \rightarrow ic_i b^\alpha_i; \ c_i^2=\left(b^\alpha_i\right)^2=\frac{1}{2}
\end{equation}
with the on-site quartic constraint
\begin{equation}\label{eq:constraint}
    D_i =  c_i b^x_ib^y_ib^z_i=\frac{1}{4}.
\end{equation}
We adapt the following unit cell convention for Majoroana fermions: each unit cell contains all $8$ Majorana fermions on sublattices A and B. They are furthermore fixed at the same position. This means that the N.N. bond vectors $\mathbf{e}_x, \mathbf{e}_y$ become the Bravais lattice vectors $\mathbf{a}_1,\mathbf{a}_2$ while $\mathbf{e}_z=1$; see Fig.~\ref{fig:lattice}(a). Thus the $\gamma_a$ in Sec.~\ref{sec:RPA} becomes naturally the collection of $c$- and $b$-Majoranas on sublattices A and B: $\gamma= (c_A, b_A^x,b_A^y,b_A^z,c_B, b_B^x,b_B^y,b_B^z)^\text{T}$.
 
\subsection{Kitaev's solution}

In the pure Kitaev limit, the Hamiltonian \eqref{eq:Kitaev-model} can be solved exactly~\cite{Kitaev2006}. After the mapping \eqref{eq:Majorana-rep},
\begin{equation}
    H = -iK\sum_{i,j \in \alpha} \hat{u}^{\alpha}_{ij}c_i c_j,
\end{equation}
where $\hat{u}^{\alpha}_{ij} = i b^\alpha_i b^\alpha_j=\pm 1/2$ are integrals of motion that commute with the constraint \eqref{eq:constraint} and the Hamiltonian. We define the gauge-invariant $\mathbb{Z}_2$ index at the center of each plaquette:
\begin{equation}\label{eq:flux}
    w_p = \prod_{i,j \in \alpha} \left(2\hat{u}^{\alpha}_{ij}\right),
\end{equation}
where the product is taken over the bonds of the plaquette. $w_p=-1$ means that there is a $\mathbb{Z}_2$ flux at the corresponding plaquette. The conservation of $\hat{u}^{\alpha}_{ij}$ then implies that the fluxes are static. A $b^\alpha$-Majorana fermion creates two fluxes in the two adjacent honeycombs with the common $\alpha$-bond, which in turn changes the sign of the N.N. $c$-Majorana coupling across that bond. Therefore, the $c$-Majorana fermion acquires a $\pi$-phase after circling around the flux which acts as a point magnetic vortex. In the ground state $|0\rangle$, there are no fluxes in the system $\hat{u}^{\alpha}_{ij} = 1/2$, and the low-energy spectrum consists of dispersive $c$-Majorana fermions.

\subsection{Majorana mean field theory}

Away from the Kitaev limit, exact solutions are not generally available. The $b$-Majoranas, and therefore the $\mathbb{Z}_2$ fluxes, become mobile which complicates their description. To obtain the single-particle spectrum we employ Majorana MFT, see e.g. Refs.~\cite{burnell20112,You2012,Ralko2014,Knolle2018,Yip2022}. For the pure Kitaev model $H_K$, this scheme reproduces the ground state spectrum of dispersive $c$-Majoranas and flat $b^\alpha$-Majorana bands; however, the $b$-Majorana band gap is not the two-flux vison gap~\cite{Knolle2018}, and the effect of $\mathbb{Z}_2$ gauge fluctuations is excluded as $c$- and $b^\alpha$-Majorana fermions are non-interacting. Here we summarize the main points, for details of the MFT see Appendix~\ref{sec:MajoranaMF}.

We decouple the Hamiltonian in the  spin-liquid channels:
\begin{equation}\label{eq:MFDecoupling-SL}
\eta_{\alpha} = i \langle c_i c_j \rangle, \ Q^{\beta \gamma}_{\alpha} =  i \langle b_i^\beta b_j^\gamma \rangle.
\end{equation}
In the above, $i, j$ belongs to sublattices A and B connected by the $\alpha$-bond. To account for the effect of an external magnetic field, we also decouple in the magnetization channel
\begin{equation}\label{eq:MFDecoupling-mfield}
 m^\alpha_i =  i \langle c_i b_i^\alpha \rangle.
\end{equation}
The magnetisation is taken to be uniform across A and B sublattices such that $m^\alpha_i$ is site-independent. We also take into account the quartic constraint \eqref{eq:constraint} by rewriting it equivalently as three quadratic constraints for the three generators of the SU$(2)$ gauge symmetry group~\cite{You2012,Ralko2014,Yip2022}
\begin{equation}
    F^\alpha_i = i \left(c_ib^\alpha_i + \frac{1}{2}\varepsilon_{\alpha\beta\gamma}b^\beta_i b^\gamma_i \right) =0.\label{eq:constraints-Majorana}
\end{equation}
The coefficients $\lambda_\alpha$ in Eq.~\eqref{eq:MF-Hamiltonian-Majorana} are then the three Lagrange multipliers, which are regarded as additional MFT parameters enforcing the constraints on average across two sublattices
\begin{equation}
    \sum_{i=A,B} \langle F^\alpha_i \rangle = 0.\label{eq:MFDecoupling-constraints}
\end{equation}

Examples of MFT spectra are presented in Appendix~\ref{sec:MFspectrum}. Here we note that $c$-Majorana spectrum remains gapless for $\mathbf{B}\parallel \mathbf{e}_b$ but becomes gapped for $\mathbf{B}\parallel \mathbf{e}_c$. $b$-Majorana bands become dispersive but remain gapped for at least small $J,\Gamma$. Finally, without external field and hence a finite magnetisation, $c$- and $b$-Majoranas are decoupled at the MFT level. Finally, we find that the MF averages retain the symmetry of the system self-consistently which is manifest in the reduced number of independent MF parameters; see Appendix~\ref{sec:MFsymmetry}.

\subsection{RPA spin susceptibility}\label{sec:RPA-spin-susceptibility}

The MFT theory of the previous subsections gives a simple non-interacting Majorana description. We now apply our general method of Sec.~\ref{sec:RPA} to include fluctuations by considering the effect of interactions. More concretely, the interaction vertex is given by Eq.~\eqref{eq:spin-Hamiltonian} written in terms of Majoranas:
\begin{equation}
    U(\mathbf{k})  = \sum_{\langle i, j \rangle \in n} O^{(n)}_{\alpha \beta} S_i^\alpha(\mathbf{k}) S_j^\beta(-\mathbf{k}) e^{-i\mathbf{k}.\mathbf{e}_{n}},
    \label{eq:interaction-vertex-spin}
\end{equation}
where $S^\alpha_i(\mathbf{k}) $ is the spin operator written in Majorana bilinears on sublattice $i= A,B$. The index $n$ in $O^{(n)}_{\alpha \beta}$ enumerates the N.N. $K, J, \Gamma$ terms and $\mathbf{e}_n$ is the corresponding bond vector. As we will see, instead of using the interaction vertex generally in the form $U_{ab,cd}$ as in Eq.~\eqref{eq:generalised-interaction-vertex}, it is more convenient to write
\begin{equation}
   U(\mathbf{k})  =  \sum_{i,j, \alpha,\beta} U_{i\alpha,j\beta}(\mathbf{k}) i c_i(\mathbf{k}- \mathbf{p}) b_i^\alpha(\mathbf{p})  i c_j(-\mathbf{k}- \mathbf{p}') b_j^\beta(\mathbf{p}').\label{eq:interaction-vertex}
\end{equation}
For the pure Kitaev limit, the non-zero matrix elements are
\begin{equation}
U_{A\alpha,B\beta}(\mathbf{k}) = \delta_{\alpha \beta} K \exp(-i \mathbf{k}.\mathbf{e}_\alpha).  \label{eq:vertex-Kitaev}
\end{equation}

Our main quantity of interest is the spin susceptibility
\begin{equation}
    \chi_{\alpha \beta}(\omega,\mathbf{k}) = \lim_{i\omega_0 \to \omega +i\delta }\sum_{i,j} \int \mathrm{d}\tau P_{\alpha \beta}(\tau,\mathbf{r}_{i} - \mathbf{r}_{j}) e^{i \omega_0 \tau -i\mathbf{k}.(\mathbf{r}_{i} -\mathbf{r}_j)}.\label{eq:spin-susceptibility-Matsubara}
\end{equation}
with the  spin-spin correlation function 
\begin{equation}
    P_{\alpha \beta}(\tau,\mathbf{r}_{i}-\mathbf{r}_j) = \langle \text{T}_\tau \{S^\alpha(\tau,\mathbf{r}_i)S^\beta(0,\mathbf{r}_j) \}\rangle.
    \label{eq:spin-correlator-Matsubara}
\end{equation}

Eqs.~\eqref{eq:spin-susceptibility-Matsubara} and \eqref{eq:spin-correlator-Matsubara} are given by Eqs.~\eqref{eq:generalised-one-loop-result} and \eqref{eq:RPA-generalised-susceptibility} in the RPA approximation. However, in our case only Majorana bilinears of the form $S_i^\alpha =ic_ib_i^\alpha$ enter in both the interactions Eq.~\eqref{eq:interaction-vertex} and the susceptibility vertices~\eqref{eq:generalised-vertex}. As a result, the explicit formulae can be considerably simplified: instead of Eq.~\eqref{eq:generalised-RPA-tensor}, one defines the RPA tensor $\chi_{i\alpha,j\beta}(\omega,\mathbf{k})$ where the sublattice and spin indices $i,\alpha$ replace $a,b$ and the same for $j,\beta$ and $c,d$. One also makes the corresponding changes in the RPA equation~\eqref{eq:generalised-RPA-tensor}; the explicit formulae are listed in Appendix~\ref{sec:KQSL-formulae}. Finally one obtains the spin susceptibility:
\begin{equation}
    \chi_{\alpha\beta}(\omega,\mathbf{k}) = \sum_{i,j= \text{A}, \text{B}}[\chi(\omega,\mathbf{k})]_{i\alpha,j\beta} e^{-i\mathbf{k}.(\mathbf{r}_i - \mathbf{r}_j)},\label{eq:RPA-susceptibility}
\end{equation}
where $\mathbf{r}_i$ and $\mathbf{r}_j$ are sublattice positions inside a unit cell. Note that here $\mathbf{r}_B -\mathbf{r}_A = -(\mathbf{a}_1+ \mathbf{a}_2)/3$ is the actual N.N. $z$-bond vector instead of unity.

At zero temperature $T=0$ and $\omega>0$, due to the fluctuation-dissipation theorem $\Im \chi_{\alpha \beta}(\omega,\mathbf{k})$ is half the dynamical form factor $S^{\alpha\beta}(\omega.\mathbf{k})$ defined in real time as:
\begin{equation}
    S^{\alpha\beta}(\omega.\mathbf{k}) = \sum_{i,j} \int_{-\infty}^{\infty} \mathrm{d}t \langle S^\alpha(t,\mathbf{r}_i)S^\beta(0,\mathbf{r}_j)\rangle e^{i\omega t -i\mathbf{k}.(\mathbf{r}_{i} -\mathbf{r}_j)},
    \label{eq:dynamical-form-factor}
\end{equation}
To make connection to experiments, we compute the intensity which is proportional to the INS cross-section:
\begin{equation}
    \mathcal{I}(\omega,\mathbf{k}) =\left(\delta_{\alpha\beta} -\frac{k_\alpha k_\beta}{k^2}\right) \Im \chi_{\alpha\beta}(\omega,\mathbf{k}). 
    \label{eq:INS-form-factor}
\end{equation}
Here $k_\alpha, k_\beta$ are 3D momenta in the global laboratory frame obtained from $\mathbf{k}$ on the $(111)$-surface using Eq.~\eqref{eq:reference-frame}.

In the remainder, unless stated otherwise, we choose $\delta = 1/L$ and $L=180$ for computing the INS intensity $\mathcal{I}$ in Eq.~\eqref{eq:INS-form-factor}. The results are plotted on a logarithmic scale $\log (1+ \mathcal{I}/20)$ to make the features of the two-Majorana continuum more visible.

\subsection{\label{sec:continuum}Two-Majorana continuum}

From the structure of the RPA we expect two different features to appear in the INS intensity. On the one hand, a broad continuum of excitations from two particle excitations similar to the Stoner continuum of a metallic magnet. On the other hand, sharper collective modes can emerge whose widths are inversely proportional to the quasi-particle life-time. Outside the two-Majorana continuum at $T=0$, these modes are  para-magnon-like with infinite life-time within RPA. Inside the continuum, these sharp spectra are strongly renormalized due to magnon decay into two Majorana fermions with the decay width determined by the imaginary part of $\chi_0$ in Fig.~\ref{fig:susceptibility-diagram}(b). As a result, broadened quasiparticle modes appear and threshold singularities at the continuum boundary. Note that the effects are similar to that of bosonic quasi-particle decay in Helium-$4$ and non-collinear magnets~\cite{Pitaevskii1959,Zhitomirsky2006}. In addition, we can have strong level repulsion between continuum and bound state excitations~\cite{rowe2012spin,verresen2019avoided}. Therefore, before presenting our results, we comment on some general features of the two-Majorana continuum in our model, which determines the region of sharp features of the collective modes.

In the RPA approximation, the two-Majorana continuum is determined by the frequency poles in the one-loop susceptibility tensor $[\chi_0(k)]_{i\alpha,j\beta}$. At zero magnetic field, the $c$- and $b$-Majorana fermions decouple in the MF approximation. Thus $\langle c b \rangle$ averages vanish identically, and $[\chi_0(k)]_{i\alpha,j\beta}$ factorizes into a product of the $c$- and $b$-Majorana Green's functions:
\begin{equation}\label{eq:one-loop-nofield}
   [\chi_0(k)]_{i\alpha,j\beta} =- \delta_{\alpha\beta} T\sum_{p_0} \sum_{\mathbf{p}} G_{ij}^c(p) G_{ji}^b(p-k);
\end{equation}
Summation over the fermion Matsubara frequency $\omega_0$ and analytical continuation $i \omega_0 \rightarrow \omega+i\delta$ gives frequency poles
\begin{equation}
    \omega \pm (E_c+E_b)+  i\delta,
\end{equation}
where $E_c$ is the $c$-Majorana energy and $E_b$ is one of the $b$-Majorana dispersions. As a result, despite  the $c$-Majorana fermions being gapless, the two-Majorana continuum could remain gapped if the $b$-Majorana fermions have a gap, which holds at small $J,\Gamma$. A finite magnetic field hybridizes the $c$- and $b$-Majorana fermions, and the dispersions $E_c$ and $E_b$ correspond to superpositions of $c$- and $b$-Majorana states. Consequently there appear additionally the poles:
\begin{equation}
    \omega \pm [E_c+E_c]+  i\delta,\  \omega \pm [E_b+E_b]+  i\delta.
\end{equation}
Thus at small  fields, the two-Majorana continuum `leaks' into energies below the $b$-Majorana gap. In particular, for magnetic field along the bond direction $\mathbf{e}_b$, $E_c$ remains gapless up to large field values~\cite{Takikawa2019,Hwang2022}, and the two-Majorana continuum is gapless near $K$-, $K'$- and $\Gamma$-points which correspond to the momentum transfer between two gapless Dirac cones.

\section{Pure Kitaev model at zero field}\label{sec:results-Kitaev}

\begin{figure}
    \centering
    \includegraphics[width=0.9\linewidth]{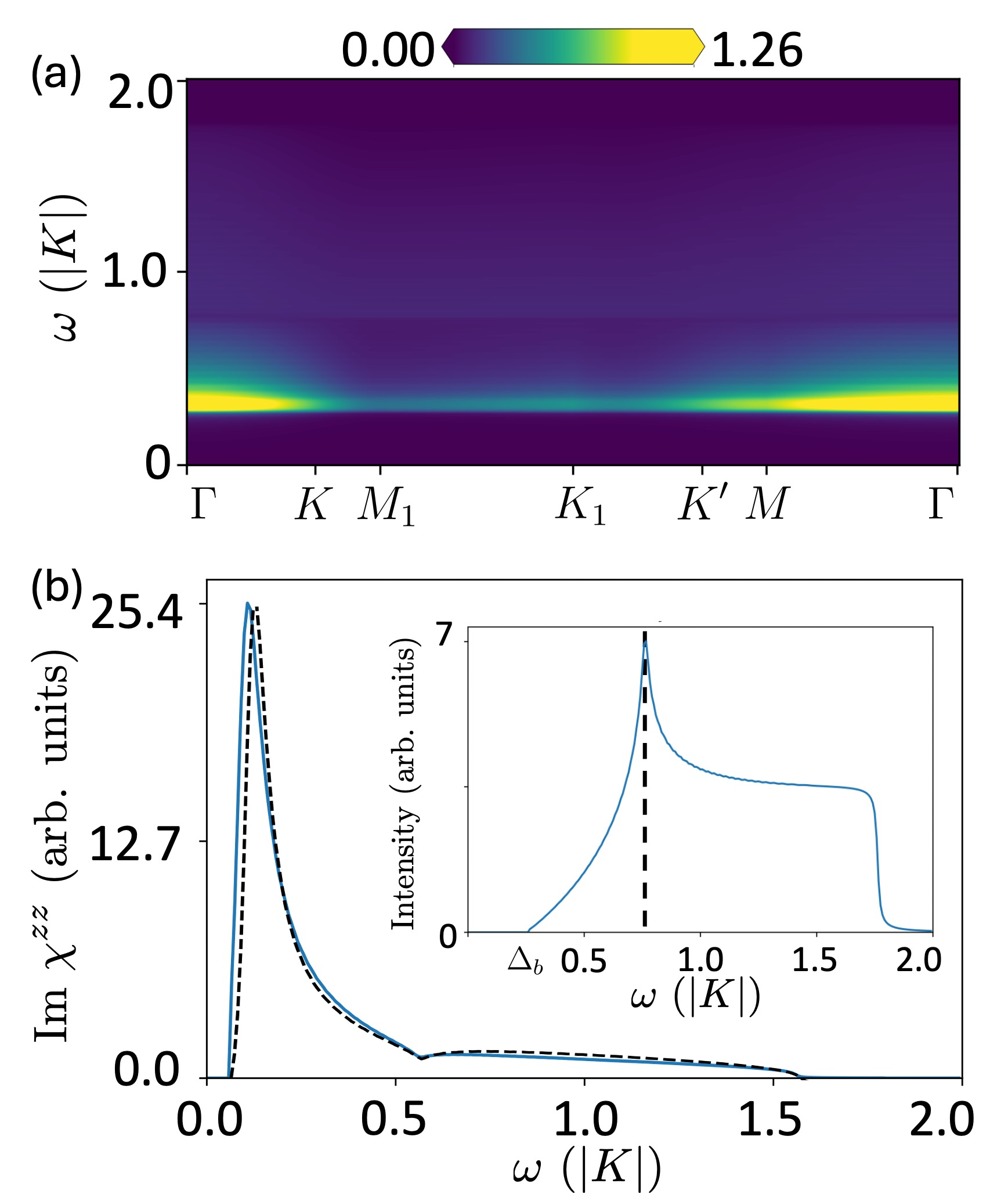}
    \caption{
The INS response of the soluble pure FM Kitaev model at zero field. (a) the INS intensity as a function of frequency and momentum path given in Fig.~\ref{fig:lattice}(c). (b) Constant momentum cut of $\Im \chi^{zz}$ at $\mathbf{k}=0$ showing the broad peak and a cusp at $\omega\approx0.57|K|$ (blue line). Dashed line: the result using the adiabatic approximation of Ref.~\cite{Knolle2014}, which is very close to the exact response but allows for a basic analytical understanding (see text). Remarkably the RPA result is almost quantitatively the same. Note, the peak intensity normalized to that of $\Im \chi^{zz}$ and the intensity is truncated at the $b$-Majorana gap $\Delta_b$ to remove unphysical broadening due to finite $\delta$. The peak onset is then shifted to the true vison gap $\Delta_v$ for comparisons. The inset is the INS intensity calculated using $\chi_0$ in Eq.~\eqref{eq:one-loop-nofield} with unshifted peak onset, showing the peak at $\omega \approx 0.76|K|$ (dashed line) due to the van-Hove singularity of the $c$-Majorana band. Fig.(b) is evaluated using $L=240$ and $\delta=1/120$ to exclude finite size effects, and plotted in linear scale.}
    \label{fig:FM-Kitaev}
\end{figure}

We now benchmark our RPA theory against the exact results of Ref.~\cite{Knolle2014}. We consider the pure FM Kitaev model at zero magnetic field with $K<0$. The resulting INS intensity is plotted in Fig.~\ref{fig:schematic}(a). It shows remarkable agreement with the exact result. In Fig.~\ref{fig:FM-Kitaev}(a) it is replotted and compared to the so-called 'adiabatic approximation'~\cite{knolle2015dynamics}. The latter is very close to the exact solution but allows for a better analytical understanding as we show in the following. 

The susceptibility contains a single peak at constant frequency across momentum regions with an off-set by the $b$-Majorana gap $\Delta_b \approx 0.26 |K|$. However, the exact two flux gap $\Delta_v \approx0.07|K|$ is a factor four smaller than the naive MFT gap of $b$-fermions as discussed and corrected in Ref.\cite{Knolle2018}. Thus in this work, we have shifted the RPA onset to the true vison gap for comparison.

In Fig.~\ref{fig:schematic}(b) a cut at $\mathbf{k}=0$ shows, in addition to the peak, a cusp at $\omega \approx 0.57 |K|$. The cusp is also present in the exact spin susceptibility~\cite{Knolle2014,knolle2015dynamics,knolle2016dynamics}, which is attributed to the van-Hove singularity at one-third-filling of the graphene-like $c$-Majorana band. 

We now elucidate analytically why our RPA theory closely resembles the almost exact result of the adiabatic approximation, which is surprising as they are based on very different physical pictures. In fact, the exact solution relies on the coupling of spin operators to a pair of local flux excitations, which provides a local scattering center for the Majorana fermions. In contrast, the RPA  does not contain local flux excitations but mimics their effect via interactions between Majoranas.

We first derive the RPA spin susceptibility for the pure Kitaev model analytically. At zero field, the one-loop susceptibility is given by Eq.~\eqref{eq:one-loop-nofield}. It is sufficient to consider only $\alpha=\beta=z$. Since $c$- and $b$-Majorana fermions decouple, instead of using the general formula it is convenient to find the $c$-Majorana Green's function directly by rewriting its Hamiltonian in terms of the doublet $\psi_\mathbf{p} = [c_A(\mathbf{p}), c_B(\mathbf{p})]^\text{T}$ as:
\begin{equation}
    H_c =\frac{1}{2} \sum_\mathbf{p}\overline{\psi}_\mathbf{p}[\Im S(\mathbf{p})\tau^x + \Re S(\mathbf{p})\tau^y ]\psi_\mathbf{p},
\end{equation}
where $S(\mathbf{p}) =(|K|/2) \sum_\alpha \exp(i\mathbf{p}.\mathbf{e}_\alpha)$, with $Q^{\alpha\alpha}_\alpha = -1/2$ in the pure Kitaev limit. The $c$-Majorana Green's function is:
\begin{align}\label{eq:c-GF-Kitaev}
    G^c(p) = - \frac{i\omega_0 +[\Im S(\mathbf{p})\tau^x + \Re S(\mathbf{p})\tau^y] }{\omega_0^2 + [E_c(\mathbf{p})]^2},
\end{align}
where $\tau^\alpha$ are the Pauli matrices in sublattice indices and the $c$-Majorana dispersion is $E_c(\mathbf{p}) =| S(\mathbf{p})|$. The fermionic Matsubara frequency is $\omega_0 = (2n+1)\pi T$. For the $b^z$-Majoranas,
\begin{align}\label{eq:b-GF-Kitaev}
    G^z(\omega_0) = -\frac{i\omega_0 -  \Delta_b \tau^y}{\omega_0^2+\Delta_b^2}.
\end{align}
Combining Eqs.~\eqref{eq:one-loop-nofield}, \eqref{eq:c-GF-Kitaev} and \eqref{eq:b-GF-Kitaev}, and summing over $\omega_0$,
\begin{align}
     &[\chi_0(\omega.\mathbf{k})]_{iz,jz} \nonumber \\
     =& - \frac{1}{2}\sum_{\mathbf{p}} \frac{E_c(\mathbf{p})+\Delta _b}{(\omega+i\delta)^2 -[E_c(\mathbf{p})+\Delta_b]^2 }
    \begin{pmatrix}1 &\cos \theta_\mathbf{p}\\\cos \theta_\mathbf{p} & 1\end{pmatrix} \nonumber \\
    =& -\frac{1}{2}\begin{pmatrix} \overline{G}(\omega) & \overline{F}(\omega)\\ \overline{F}(\omega)& \overline{G}(\omega) \end{pmatrix}.
\label{eq:one-loop-pureKitaev}
\end{align}
$\theta_\mathbf{p} = -\arctan [\Im S(\mathbf{p})/\Re S(\mathbf{p})]$ is odd in momentum. The one-loop susceptibility does not depend on $\mathbf{k}$ because the b-Majoranas are purely local as seen from their flat dispersion. We can then compute the RPA susceptibility component $\chi^{zz}(\omega,\mathbf{k})$ at $\mathbf{k}=0$. In the pure Kitaev limit, the interaction vertex \eqref{eq:vertex-Kitaev} conserves the spin index $\alpha$. Therefore the summation of one-loop diagrams in Fig.~\eqref{fig:susceptibility-diagram}(c) is independent for each diagonal spin component with $[\hat{U}(\mathbf{k})]_{iz,jz} = -|K| (\tau^x)_{ij}$, and the RPA equation \eqref{eq:generalised-RPA-tensor} reduces to a matrix equation in sublattice indices only. Then, the RPA spin susceptibility has the structure $\chi^{\alpha\beta} \propto \delta_{\alpha\beta}$. At $\mathbf{k}=0$, Eq.~\eqref{eq:RPA-susceptibility} becomes
\begin{equation}
\begin{split}
    \chi^{zz}(\omega,0) =& \chi^{zz}_{AA}(\omega) + \chi^{zz}_{BB}(\omega)+ \chi^{zz}_{BA}(\omega)+ \chi^{zz}_{AB}(\omega)   \\
    =&-\frac{ \overline{G}(\omega)+\overline{F}(\omega)}{1+(|K|/2)[\overline{G}(\omega)+\overline{F}(\omega)]}.
\end{split}\label{eq:RPA-zz-Kitaev}
\end{equation}

We now compare Eq.~\eqref{eq:RPA-zz-Kitaev} with analytical results from the adiabatic approximation~\cite{Knolle2014}. We briefly summarize the physical idea behind the approach which takes into account the $\mathbb{Z}_2$ flux excitations explicitly as an X-ray edge problem. For the rest of this subsection we use the zero temperature real-time Green's functions which are more suitable for our purpose.

\begin{figure}
    \centering
    \includegraphics[width=\linewidth]{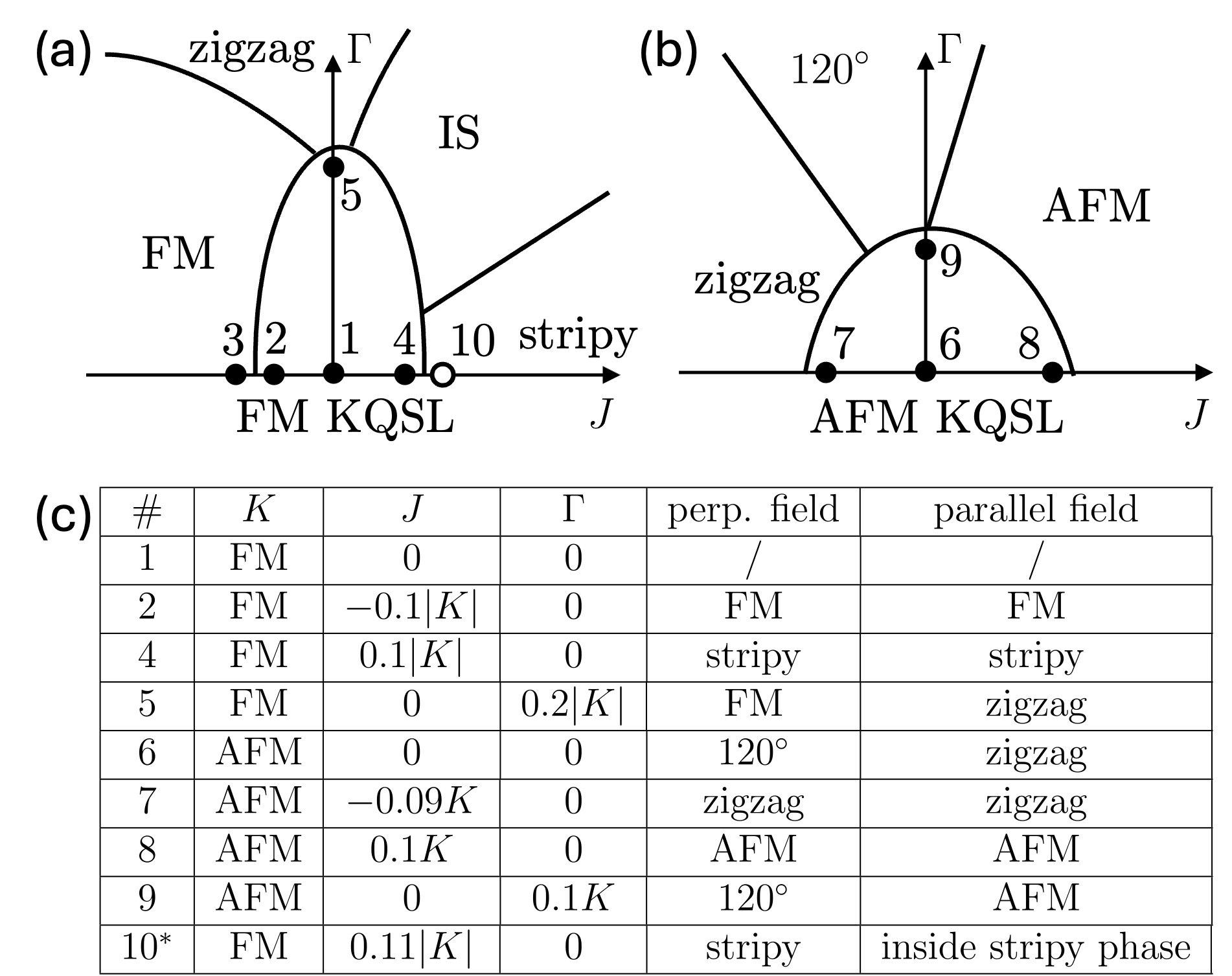}
    
    \caption{
    Schematic phase diagrams of the $KJ\Gamma$-model at zero field near (a) the pure  Kitaev FM (b) AFM Kitaev. The phase diagrams are adapted from Ref.~\cite{Rau2014}. The parameter values considered in this paper are marked by circles. (c) Summary of instabilities induced by finite field for the parameters in (a) and (b). One realization for the `field-induced' KQSL is given by parameters marked by the white circle in (a) and by $*$ in (c).}
    \label{fig:phase-schematic}
\end{figure}

\begin{figure*}
    \centering
    \includegraphics[width=\linewidth]{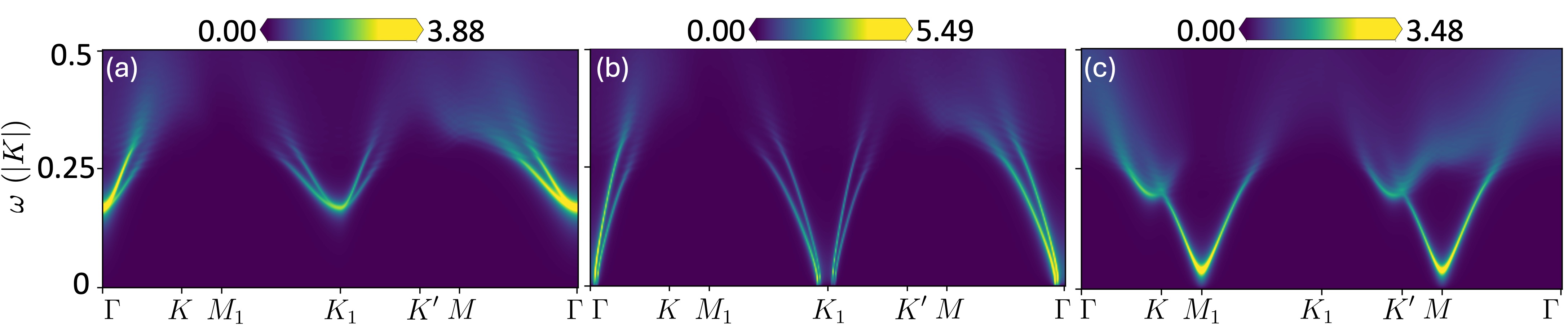}
    \caption{
    The RPA INS intensity of the $KJ\Gamma$-model at zero magnetic field with FM Kitaev coupling $K<0$ for: (a) $J=-0.1|K|, \Gamma=0$; (b) $J=-0.13|K|, \Gamma=0$; and (c) $J=0.1|K|, \Gamma=0$.  The FM Heisenberg coupling induces a sharp mode at the $\Gamma$-point which condenses at sufficiently large $|J|$. For AFM Heisenberg coupling, sharp modes appear at $M$- and $M_1$-points, signifying instability towards the stripy phase. 
    (b) signifies a FM instability due to increased $|J|$.
    }
    \label{fig:FM-Kitaev-zero-field-J}
\end{figure*}

\begin{figure*}
    \centering
    \includegraphics[width=\linewidth]{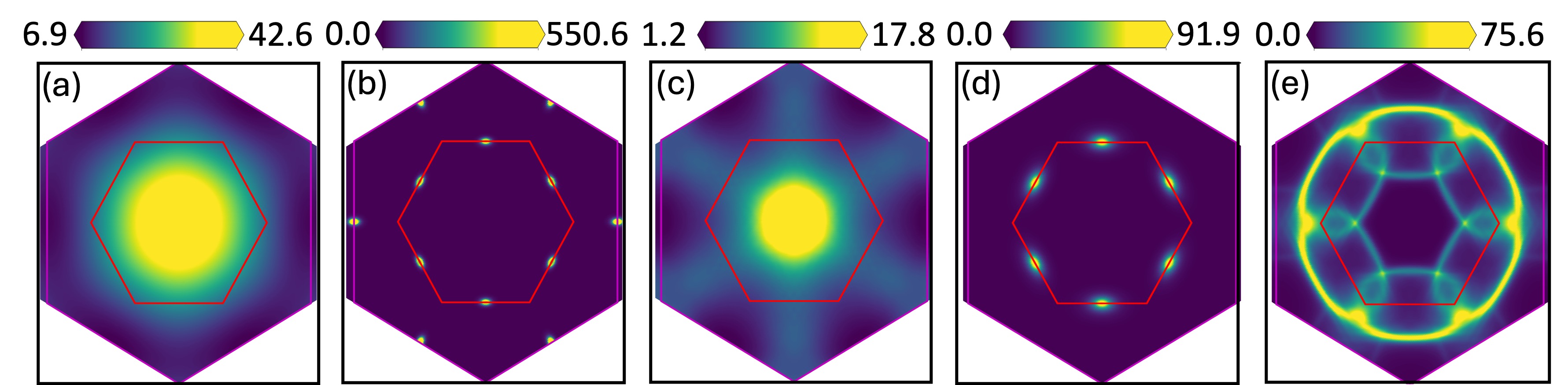}
    \caption{
    Constant frequency cuts of RPA INS intensities at zero magnetic field for: (a) the pure FM Kitaev model at $\omega = 0.3 |K|$; FM Kitaev coupling and $J=0.1|K|, \Gamma=0$ at (b) $\omega=0.045 |K|$ and (c) $\omega=0.45 |K|$; AFM Kitaev coupling and $J=-0.09K$ at (d) $\omega=0.1 K$ and (e) $\omega=0.25K$. The first and second BZs are shown in red and magenta. All figures are plotted in linear scale.}
    \label{fig:zero-field-const}
\end{figure*}

Consider the dynamical form factor Eq.~\eqref{eq:dynamical-form-factor} in real space:
\begin{equation}
    S^{\alpha\beta}_{ij}(t; \mathbf{r}_1,\mathbf{r}_2) = \langle 0|i c_i(t,\mathbf{r}_1) b_i^\alpha(t,\mathbf{r}_1) i c_j(0,\mathbf{r}_2)b_j^\beta(0,\mathbf{r}_2)|0\rangle. \label{eq:dynamical-form-factor-1}
\end{equation}
Eq.~\eqref{eq:dynamical-form-factor-1} has the physical meaning that at time $t=0$, the $b^\beta$-Majorana  creates from the zero-flux ground state $|0\rangle$ a pair of $\mathbb{Z}_2$ fluxes in the two honeycombs sharing the bond $\alpha$. The $c$-Majorana fermions then propagate in the field of the two fluxes until they are annihilated at later time $t$ by $b^\alpha$, since the system must return to $|0\rangle$. Accordingly only on-site and N.N. real-time spin-spin correlation functions are non-zero:
\begin{equation}
    S^{\alpha\beta}_{ii}(t; \mathbf{r},\mathbf{r})  =  \delta_{\alpha \beta}S^{\alpha \alpha}_{ii}(t); \ S^{\alpha\beta}_{AB}(t; \mathbf{r},\mathbf{r}+\mathbf{e_\alpha})  = \delta_{\alpha \beta}S^{\alpha \alpha}_{AB}(t). \label{eq:dynamical-form-factor-2}
\end{equation}
It can be shown that the spin susceptibility is
\begin{equation}
    \Im \chi^{\alpha \beta}(\omega,\mathbf{k}) =\delta_{\alpha \beta}\left[ S^{\alpha \alpha}_{AA}(\omega) + S^{\alpha\alpha}_{AB}(\omega) \cos (\mathbf{k}.\mathbf{e}_\alpha) \right].
\end{equation}
We consider only $\alpha=\beta=z$ and take $b^z$ in Eqs.~\eqref{eq:dynamical-form-factor-1} and \eqref{eq:dynamical-form-factor-2} to be at the origin (remember $\mathbf{e}_z =0$). Here the quantities of interest are the complex fermions:
\begin{equation}
    f(t,\mathbf{r}) = \frac{1}{\sqrt{2}}\left[c_A(t,\mathbf{r}) +i c_B(t,\mathbf{r})\right],
\end{equation}
and the time-ordered Green's function at the origin reads
\begin{equation}
    G_f(\omega) = -i \int\mathrm{d}t \ e^{i\omega t}  \langle 0|\text{T} \{ f(t,0) \hat{S}(t,0)f^\dagger(0,0) \}|0 \rangle.
\end{equation}
The scattering operator
\begin{equation}
    \hat{S}(t_1,t_2) =  e^{-i\int_{t_2}^{t_1} V(t)\mathrm{d}t}, \ V(t) = |K|\left[f^\dagger(t,0) f(t,0) -\frac{1}{2} \right],
\end{equation}
describes the potential $V$ of  fluxes created by $b^z$,  changing the sign of the $c$-Majorana coupling along the $z$-bond. For $\omega>0$,
\begin{equation}
    \chi^{zz}(\omega,0) = -G_f(\omega),
\end{equation}
Thus finding $\chi^{zz}$ is formally equivalent to the X-ray edge problem, and the adiabatic approximation consists in regarding the potential as switching on and off adiabatically: $S(t,0) \approx S(\infty,-\infty)$, whence $G_f(\omega)$ admits the simple solution:
\begin{equation}
G_f(\omega) = \frac{\overline{G}_0(\omega)}{1+ |K| \overline{G}_0(\omega)}, \ \overline{G}_0(\omega) = \sum_\mathbf{p} G_0(\omega,\mathbf{p}).
\end{equation}
The free Green's function $G_0(\omega,\mathbf{p})$ is given by
\begin{equation}
    G_0(\omega,\mathbf{p}) = \frac{\cos^2 (\theta_{\mathbf{p}}/2)}{\omega - E_c(\mathbf{p})+i\delta. \omega}+ \frac{\sin^2 (\theta_{\mathbf{p}}/2)}{\omega + E_c(\mathbf{p}) + i\delta. \omega};
\end{equation}
Note our definition of $\theta_{\mathbf{p}}$ is twice that of Ref.~\cite{Knolle2014}. Thus,  the spin susceptibility in the adiabatic approximation is given by
\begin{subequations}\label{eq:adiabatic-Kitaev}
\begin{align}
    \chi^{zz}_{\text{ad}}(\omega,0) &=  -\frac{\overline{F}_1(\omega)+\overline{G}_1(\omega)  }{1+|K|[ \overline{F}_1(\omega)+\overline{G}_1(\omega) ]},\\
   \overline{G}_1(\omega)& = \sum_{\mathbf{p}} \frac{ \omega}{(\omega+i\delta. \omega)^2 -[E_c(\mathbf{p})]^2 }, \\
   \overline{F}_1(\omega)& = \sum_{\mathbf{p}} \frac{ E_c(\mathbf{p})\cos \theta_{\mathbf{p}}}{(\omega+i\delta. \omega)^2 -[E_c(\mathbf{p})]^2}.
\end{align}
\end{subequations}

Eq.~\eqref{eq:adiabatic-Kitaev} and the RPA result Eq.~\eqref{eq:RPA-zz-Kitaev} indeed are very similar! We plot them together in Fig.~\ref{fig:FM-Kitaev}(b). The pole structure of the impurity potential given by the local flux excitation of the exact solution has been replaced by the interaction within RPA. However, we find $|K|$ instead of $|K|/2$ enters Eq.~\eqref{eq:adiabatic-Kitaev}, and the definitions of the functions $\overline{G},\overline{F}$ and $\overline{G}_1,\overline{F}_1$ are slightly different. However, for both definitions the integration over $\mathbf{p}$ diverges logarithmically at frequency $\omega = 0.5 |K|$ [offset by $\Delta_b$ in Eq.~\eqref{eq:RPA-zz-Kitaev}], which is due to the logarithmic divergence of $c$-Majorana density of states at the van-Hove point explaining the cusp in Fig.~\ref{fig:FM-Kitaev}(b) at $\omega = 0.5 |K|+\Delta_v \approx 0.57 |K|$ (remember we have shifted the peak onset from $\Delta_b$ to $\Delta_v$). In Fig.~\ref{fig:FM-Kitaev}(b) inset, we also show the non-interacting INS intensity which contains a similar unshifted divergence at $\omega = 0.5|K| + \Delta_b \approx 0.76 |K|$.

\section{$KJ\Gamma$-model at zero field -- appearance of sharp modes}\label{sec:results-KJGamma}

\begin{figure}
    \centering
    \includegraphics[width=0.9\linewidth]{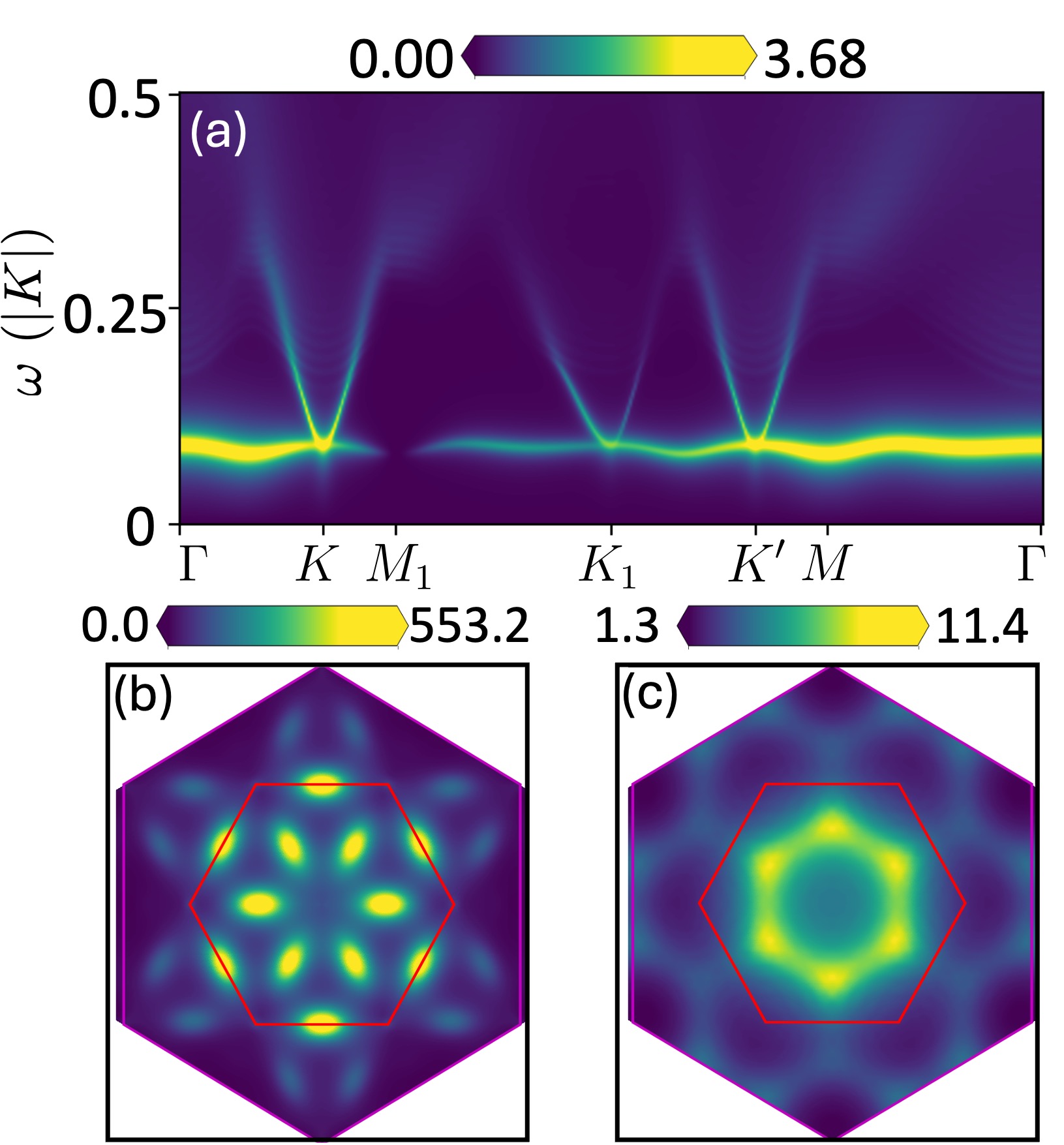}
    \caption{
    The RPA INS intensity of the $KJ\Gamma$-model at zero magnetic field with FM Kitaev coupling $K<0$ at $J=0, \Gamma=0.2|K|$. (a) The INS intensity as a function of frequency along the BZ path in Fig.~\ref{fig:lattice}(c). There appear a continuous sharp mode at $M$-points and across broad regions of the BZ are attributed to the system's simultaneous affinity to the zig-zag and incommesurate spiral (IS) phases. (b) shows a constant frequency cut at $\omega \approx 0.08|K|$ in linear scale intersecting the broad mode minima corresponding to the two instabilities.  (c) constant frequency cut at $\omega \approx 0.45|K|$. 
    }
    \label{fig:FM-Kitaev-zero-field-Gamma}
\end{figure}

\begin{figure*}
    \centering
    \includegraphics[width=\linewidth]{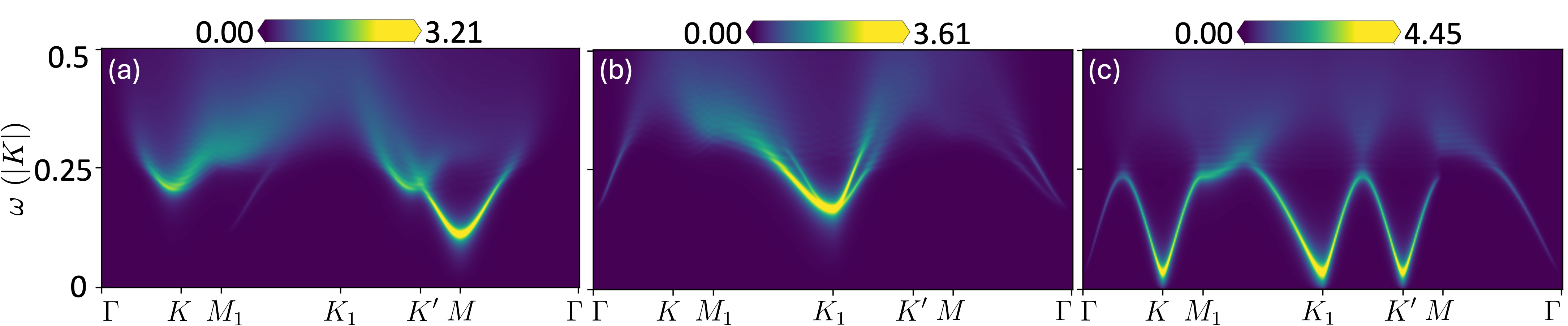}
    \caption{
    The RPA INS intensity for the $KJ\Gamma$-model at zero magnetic field with AFM Kitaev coupling $K>0$ and: (a) $J=-0.09K, \Gamma=0$; (b) $J=0.1K, \Gamma=0$; (c) $J=0, \Gamma=0.1K$. The $M$- and $K_1$-modes in (a) and (b) signify zig-zag and AFM phase instabilities respectively. In (c) the instabilities are towards the $120^\circ$ ($K$-modes) and the AFM ($K_1$-modes) phases. 
    }
    \label{fig:AFM-Kitaev-zero-field}
\end{figure*}

Away from the pure Kitaev model, we compute the INS intensity Eq.~\eqref{eq:INS-form-factor} for both ferro- and antiferromagnetic Kitaev couplings for various values of $ J,\Gamma$. As one of the central qualitative results, we find that whithin the KQSL the additional Majorana interactions result in {\it sharp} para-magnon modes, which coexist with the Majorana continuum. Shifting these in energy the additional interaction terms induce instabilities towards magnetically ordered phases. 

They are highly sensitive to the strength of $J, \Gamma$ condensing already small values [below we show one such example in Fig.~\ref{fig:FM-Kitaev-zero-field-J}(b)]. Within the RPA, this  indicates instabilities towards magnetic ordering of the KQSL due to $J, \Gamma$ couplings. The momenta of the different magnetic orders are shown in Fig.~\ref{fig:lattice}(c). Another key result of our work is that the location of the mode condensation in the BZ  coincides with those of magnetic orders found in ED studies of the phase diagram in Ref.~\cite{Rau2014}, as summarized schematically in Fig.~\ref{fig:phase-schematic}(a)-(b). In the rest of this section we shall present our results in detail.

\subsection{FM Kitaev coupling}
We first consider two cases with FM Kitaev coupling $K<0$. They are marked $2, 3$ in the schematic phase diagram Fig.~\ref{fig:phase-schematic}(a).
A small FM Heisenberg coupling $J=-0.1 |K|$ immediately induces a sharp mode near $\mathbf{k}=0$ in the INS intensity; see Fig.~\ref{fig:FM-Kitaev-zero-field-J}(a). There is additionally a much weaker mode at $K_1$-points due to periodicity of the BZ. However, at slightly higher FM Heisenberg coupling strength $J=-0.13 |K|$ the sharp mode condenses, as seen in Fig.~\ref{fig:FM-Kitaev-zero-field-J}(b). This means the KQSL is unstable with respect to interactions and the ground state is a ferromagnet, in qualitative agreement with ED~\cite{Rau2014} and DMRG~\cite{Gohlke2017} results. We verify that the critical $J_{\text{cr}}$ value lies between $-0.12|K|$ and $-0.13 |K|$, which agrees well with the DMRG result $J_{\text{cr}} \approx -0.116 |K|$~\cite{Gohlke2017}.

An AFM Heisenberg coupling $J=0.1 |K|$ results in sharp modes at both $M$- and $M_1$-point at zero field, as shown in Fig.~\ref{fig:FM-Kitaev-zero-field-J}(c). This indicates an instability into a stripy phase in agreement with ED~\cite{Rau2014} and DMRG~\cite{Gohlke2017} results; cf. position $4$ in Fig.~\ref{fig:phase-schematic}(a). Upon small increase to $J=0.11|K|$, the sharp modes condense at zero field which means the system is in the stripy phase; see Fig.~\ref{fig:proximate-KQSL}(a) below. For reference, in DMRG $J_{\text{cr}}\approx 0.094 |K|$~\cite{Gohlke2017}.

Here we note that despite the presence of sharp modes at lower energy, at higher frequencies the INS intensities contain broad regions of a featureless continuum, similar to the pure Kitaev model. In Fig.~\ref{fig:zero-field-const}(a)-(c) we plot the constant frequency cuts of INS intensities for the FM Kitaev model and for $J=0.1 |K|$. For the FM Kitaev model, the cut is self-similar at all energies, since there is only one energy scale given by the Kitaev coupling. For $J=0.1 |K|$ the low-frequency cut at $\omega = 0.045 |K|$ shows minima of the sharp, $C_6$-symmetric $M$- and $M_1$-modes. At higher frequency $\omega = 0.45 |K|$ the INS intensity is featureless and resembles that of the pure FM Kitaev model.

\begin{figure*}
    \centering
    \includegraphics[width=\linewidth]{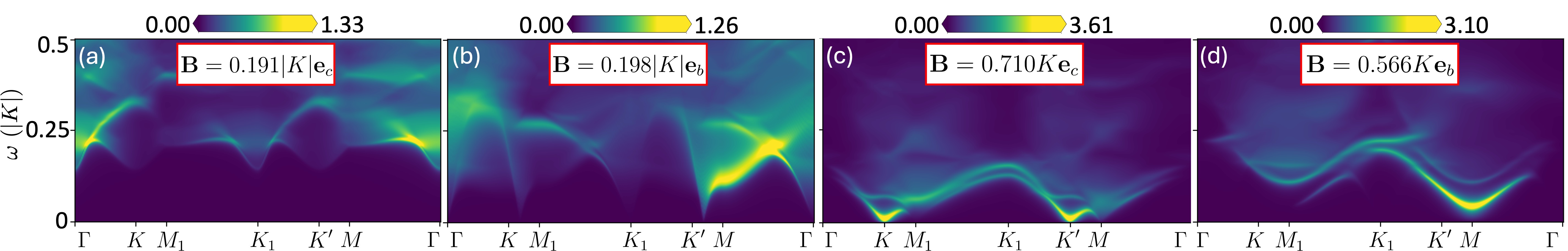}
    \caption{
    The RPA INS intensity for the pure FM and AFM Kitaev model at finite external magnetic fields $\mathbf{B}$. (a) FM $K<0$ and $\mathbf{B} = 0.191 |K|\mathbf{e}_c$ perpendicular to the honeycomb lattice. (b) FM $K<0$ and $\mathbf{B} = 0.198 |K|\mathbf{e}_b$ along the bond direction. The magnetic fields for (a) and (b) are at MF critical values. (c) AFM $K>0$ and $\mathbf{B} = 0.710 K \mathbf{e}_c$. There is a sharp mode near $K$-points; upon increasing the field, this mode condenses, signifying a phase transition into a $120^\circ$ phase. (d) AFM $K>0$ and $\mathbf{B} = 0.566K\mathbf{e}_b$. The sharp mode is located at the $M$-point, which also condenses as field strength increases. The system then enters the zigzag-phase.
    }
    \label{fig:pure-Kitaev-field}
\end{figure*}

\begin{figure*}
    \centering
    \includegraphics[width=\linewidth]{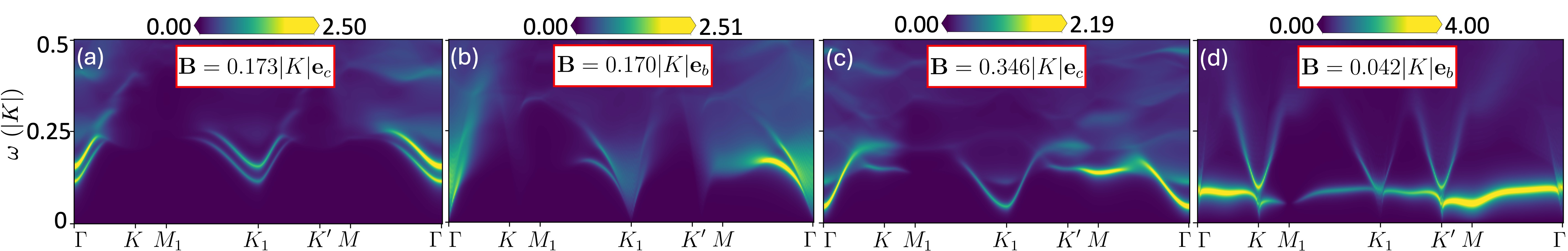}
    \caption{
    The RPA INS intensity for the $KJ\Gamma$-model with FM Kitaev coupling $K<0$. (a)-(b): $J=-0.1|K|, \Gamma=0$ at (a) $\mathbf{B}=0.173|K|\mathbf{e}_c$ and (b) $\mathbf{B}=0.170|K|\mathbf{e}_b$. In (a)-(b), the field values are at the MF critical points. In all cases, the FM Heisenberg term induces a sharp mode near the $\Gamma$-point. The mode decreases in energy at finite field. In (b) the mode is `pinched' by the gapless two-Majorana continuum near the $\Gamma$-point. (c)-(d): $J=0, \Gamma=0.2|K|$ at (c) $\mathbf{B}=0.346|K|\mathbf{e}_c$ and (d) $\mathbf{B}=0.042|K|\mathbf{e}_c$. In (c) the field is at the MF critical point along $\mathbf{e}_c$, and the FM mode at the $\Gamma$-point resurfaces. The system is much less stable along $\mathbf{e}_b$, which favours the zigzag $M$-mode, as shown in (d). Upon a small further increase of the magnetic field, the $M$-mode immediately condenses. Note that the sharp mode is repelled away from $K$-points by the gapless two-Majorana continuum.
    }
    \label{fig:FM-Kitaev-field}
\end{figure*}

We also consider the case with AFM $\Gamma=0.2|K|$ [position $5$ in Fig.~\ref{fig:phase-schematic}(a)], which is physically relevant to the Kitaev candidate material $\alpha$-RuCl$_3$~\cite{Maksimov2020}. We find sharp peaks in the INS intensity across broad regions of the BZ as shown in Fig.~\ref{fig:FM-Kitaev-zero-field-Gamma}(a). In Fig.~\ref{fig:FM-Kitaev-zero-field-Gamma}(b) we plot the constant frequency cut at $\omega \approx 0.08|K|$ which is the peak minimum. We find these peak minima to be at $M$-points as well as lower-symmetry points within the first BZ. Here according to the ED result~\cite{Rau2014}, the spin Hamiltonian \eqref{eq:spin-Hamiltonian} ground state is close to FM, zigzag and incommensurate spiral (IS) phases. We therefore attribute the sharp peaks at non-symmetric wave-vectors to the IS instability, whose ordering wave vectors do not lie at the BZ high-symmetry points; see Fig.~\ref{fig:FM-Kitaev-zero-field-Gamma}(b). In fact, the incommensurate-mode locations agree qualitatively with the ordering wave-vector of the IS phase from ED studies~\cite{Rau2014}. In all cases considered, the existing sharp modes indicate magnetic instabilities in qualitative agreement with ED studies. In Fig.~\ref{fig:FM-Kitaev-zero-field-Gamma}(c) we show the cut at higher frequencies $\omega =0.45 |K|$ which also shows the featureless continuum. 

\subsection{AFM Kitaev coupling}

Next, we turn to AFM Kitaev couplings $K>0$. 
The case with FM Heisenberg coupling $J=-0.09 K$ at zero field is plotted in Fig.~\ref{fig:AFM-Kitaev-zero-field}(a) [marked as $7$ in Fig.~\ref{fig:phase-schematic}(a)]. There a sharp mode appears at $M$-points in agreement with the ED result that the system is in a zigzag order~\cite{Rau2014}. We also show constant-frequency cuts in Fig.~\ref{fig:zero-field-const}(d)-(e). At low frequency $\omega = 0.1 K$, sharp modes appear near $M$-points whereas at higher frequency $\omega = 0.25 K$, the INS intensity is concentrated around the first BZ boundary. These features resemble DMRG simulations of spin susceptibility in similar parameter regimes in the zigzag-phase~\cite{Gohlke2017}.

The AFM Heisenberg coupling $J=0.1 K$ at zero magnetic field induces sharp modes at $K_1$-points on the second BZ, as shown in Fig.~\ref{fig:AFM-Kitaev-zero-field}(b). As expected this corresponds to an AFM instability; see position $8$ in Fig.~\ref{fig:phase-schematic}(a). Here we note that the DMRG critical $J$ values at which the KQSL becomes unstable are $J_{\text{cr}} \approx \pm 0.020 |K|$~\cite{Gohlke2017}. Therefore the RPA critical $J$ values are significantly larger in contrast to the FM Kitaev case in the previous section. Overall, the RPA overestimates the stability of KQSL. This is expected as RPA neglects the part of the $\mathbb{Z}_2$ gauge field fluctuations. Concretely, it captures the influence of vison pairs, which are directly related to the $b$-fermion modes, but cannot capture individual visons.

Finally, we consider a small AFM $\Gamma =0.1 K$ which gives sharp modes both at $K$- and $K_1$-points; see Fig.~\ref{fig:AFM-Kitaev-zero-field}(c). This is qualitatively in agreement with Ref.~\cite{Rau2014} where the system lies on the boundary between AFM and $120^\circ$ phases, and also shown schematically at position $9$ in the schematic phase diagram Fig.~\ref{fig:phase-schematic}(a).

\begin{figure}
    \centering
    \includegraphics[width=0.9\linewidth]{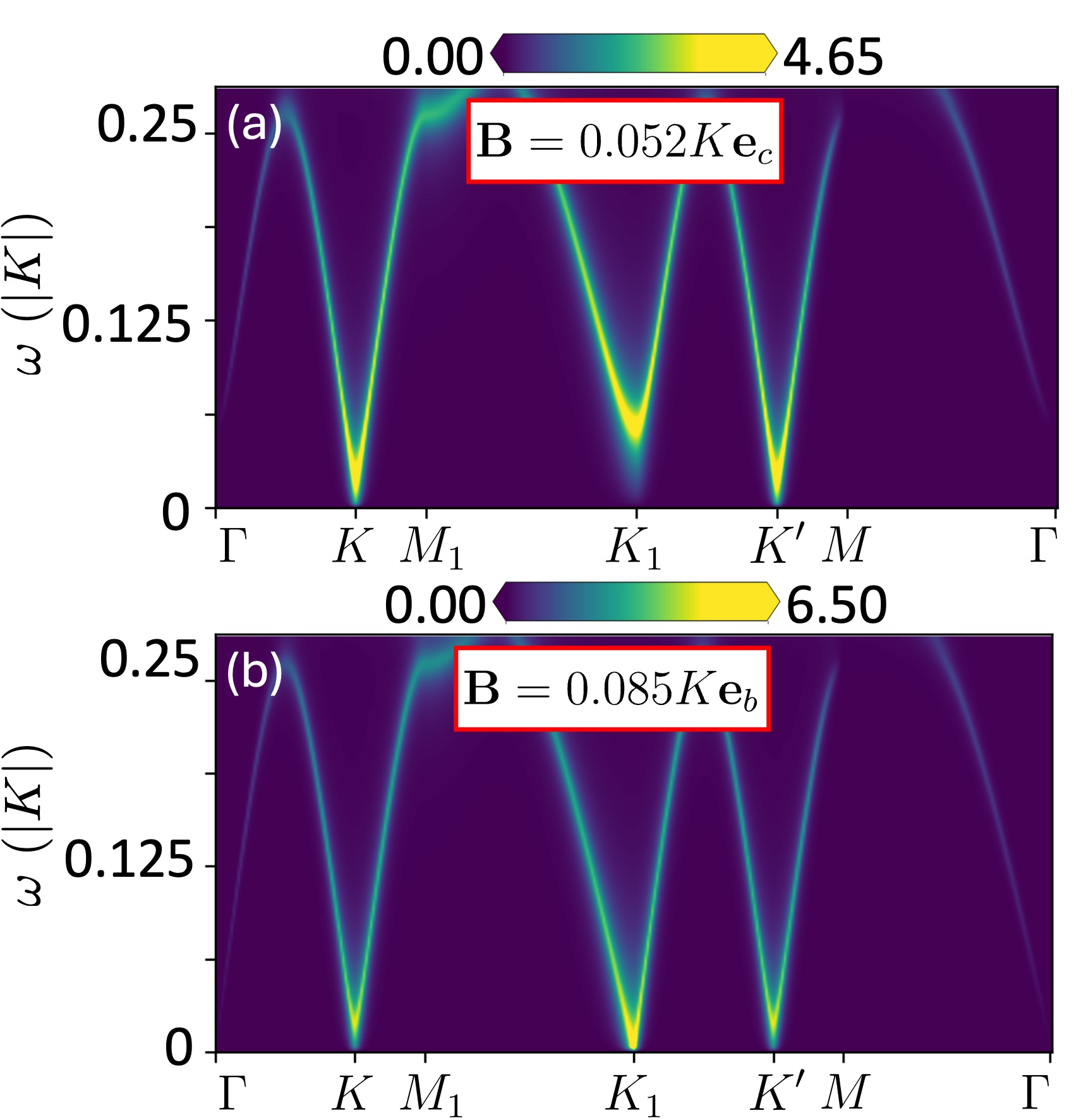}
    \caption{
    The RPA INS intensity for the $KJ\Gamma$-model with AFM Kitaev coupling $K>0$ at $J=0, \Gamma=0.1K$ and: (a) $\mathbf{B}=0.052|K|\mathbf{e}_c$ perpendicular to the lattice; (b) $\mathbf{B}=0.085|K|\mathbf{e}_b$ along the bond direction. For $\mathbf{B}\parallel \mathbf{e}_c$, the $K_1$-mode is lifted whereas the $K$-mode is lowered. For $\mathbf{B}\parallel \mathbf{e}_b$, both modes are lowered but the $K_1$-mode condenses first at $|\mathbf{B}|=0.085K$. Note the KQSL becomes unstable at small field values for both field directions.
    }
    \label{fig:AFM-Kitaev-field}
\end{figure}

\section{Finite magnetic field}\label{sec:results-finite-field}

We next consider the effect of external magnetic field $\mathbf{B}$ along $\mathbf{e}_c$ and $\mathbf{e}_b$ directions for the parameters considered in the previous sections. The results are summarized in the table in Fig.~\ref{fig:phase-schematic}(c). We show the INS intensities for some of the parameter values there below.

\subsection{Pure Kitaev model}
At finite magnetic field, the FM and AFM Kitaev models exhibit different behaviors. We present the INS intensities in Fig.~\ref{fig:pure-Kitaev-field}(a)-(d). For the FM Kitaev model, the KQSL remains robust up to the critical field values of $|\mathbf{B}_{\text{cr}}| \approx 0.191 |K|$ for $\mathbf{B}\parallel \mathbf{e}_c$ in agreement with Ref.~\cite{Yip2022}, and $|\mathbf{B}_{\text{cr}}| \approx 0.198 |K|$ for $\mathbf{B}\parallel \mathbf{e}_b$ respectively. The corresponding INS intensities are shown in Fig.~\ref{fig:pure-Kitaev-field}(a)-(b). In both cases, no sharp modes develop until the critical point. The single magnetic transition has been previously observed for $\mathbf{e}_c$ direction in DMRG simulations~\cite{Gohlke2018}. For $\mathbf{B}\parallel \mathbf{e}_c$, the INS intensity is gapped, due to the field-induced $c$-Majorana gap. For $\mathbf{B}\parallel \mathbf{e}_b$, a broad, yet visible peak develops across the first BZ on the background of gapless two-Majorana continuum, which arises since the $c$-Majorana fermions remain gapless.

For AFM Kitaev coupling, the MF critical fields are much larger at $|\mathbf{B}_{\text{cr}}| \approx 0.814K$ for $\mathbf{B}\parallel \mathbf{e}_c$ in agreement with Ref.~\cite{Yip2022}, and $|\mathbf{B}_{\text{cr}}| \approx 0.778K$ for $\mathbf{B}\parallel \mathbf{e}_b$. The KQSL MFT state becomes unstable with respect to interactions already at lower fields. This happens at $\mathbf{B} = 0.710 K\mathbf{e}_c$, where the magnon mode are about to condense at $K$-points [Fig.~\ref{fig:AFM-Kitaev-field}(c)] indicating a transition to the $120^\circ$ phase. However, we do not see  successive topological transitions for increasing magnetic field  as observed in DMRG simulations~\cite{Gohlke2018}. This may be due to the uniform magnetization ansazt for $\mathbf{m}$ and the two-site unit cell assumption, which is a bad approximation for $K>0$ at small fields. For $\mathbf{B} = 0.57 K\mathbf{e}_b$, the mode is near condensation at the $M$-points [Fig.~\ref{fig:AFM-Kitaev-field}(d)], signifying the transition into a zigzag phase. Note that for both cases, the field values considered exceed the regime of validity of the perturbation theory. 

In general, the effect of a magnetic field is much richer for AFM Kitaev couplings. The weakening of the KQSL state by external magnetic fields can be understood as follows: the magnetic field compresses the MF Majorana bandwidth and shifts it to lower energies. Therefore, the sharp modes, which develop from the two-Majorana continuum, condense more easily, triggering transitions to long range ordered states. 

\begin{figure*}
    \centering
    \includegraphics[width=\linewidth]{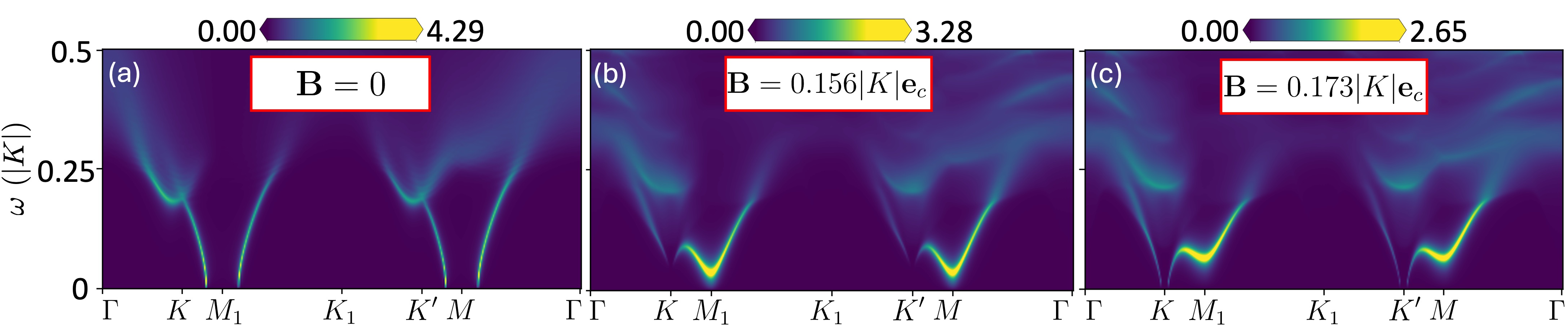}
    \caption{
Field-induced KQSL scenario with FM $K<0$ and $J=0.11|K|, \Gamma = 0$. At zero field the system is in a stripy phase, since the modes at $M$- and $M_1$-points are condensed, see panel (a). These unstable modes are lifted to finite frequency as the external field increases shown in panels (b) and (c), stabilizing the KQSL. }
    \label{fig:proximate-KQSL}
\end{figure*}

\subsection{$KJ\Gamma$-model}

The effect of finite magnetic field is generally similar to the pure Kitaev case, i.e.\ the magnetic field tends to decrease the Majorana bandwidth and destabilizes the MF KQSL into one of the competing magnetically ordered phases via the condensation of sharp modes at the corresponding BZ points. The KQSL is, however, more robust under perpendicular fields, which might be due to the induced $c$-Majorana gap. This possibility is demonstrated in the schematic phase diagram of Fig.~\ref{fig:schematic}(b). The INS results are presented in Figs.~\ref{fig:FM-Kitaev-field} and \ref{fig:AFM-Kitaev-field}.

In Fig.~\ref{fig:FM-Kitaev-field}(a)-(b), we show the results for FM $K$ and $J=-0.1|K|, \Gamma=0$. Compared to the zero field case in Fig.~\ref{fig:FM-Kitaev-zero-field-J}(a), the $\Gamma$-mode becomes damped with increasing $\mathbf{B}\parallel\mathbf{e}_c$. As the field increases along the $\mathbf{e}_b$-direction, the mode is lowered and becomes `pinched' by the gapless two-Majorana continuum near the $\Gamma$-point, as discussed in Sec.~\ref{sec:continuum}. For both directions of the field, the magnon mode does not condense up to $\mathbf{B}_{\text{cr}}$. 

Fig.~\ref{fig:FM-Kitaev-field}(c)-(d) demonstrate the results for FM $K$ and $J=0, \Gamma=0.2|K|$; the zero field results are given in Fig.~\ref{fig:FM-Kitaev-zero-field-J}(c). A finite magnetic field along $\mathbf{e}_c$ produces a sharp $\Gamma$-mode at $|\mathbf{B}_{\text{cr}}|=0.346|K|$ corresponding to FM correlations, but the MF KQSL is stable up to the MF critical field. This is shown in Fig.~\ref{fig:FM-Kitaev-field}(c). In contrast, the $\mathbf{e}_b$ direction field instead causes a sharp $M$-mode to condense already at much smaller field $\mathbf{B}\sim 0.05|K|\mathbf{e}_b$; see Fig.~\ref{fig:FM-Kitaev-field}(d). Here we also see the level repulsion from $K$-points by the gapless two-Majorana continuum as discussed in Sec.~\ref{sec:continuum}.

For AFM Kitaev coupling and $J=-0.09K, 0.1 K$, the given modes shown in Fig.~\ref{fig:AFM-Kitaev-zero-field}(a)-(b) condense before the MF critical field values for both field directions (not shown here). This means that for $J=-0.09K$, the system transitions into a zigzag-phase with increasing magnetic field, whereas for $J=0.1K$, a finite magnetic field turns the KQSL into a conventional N\'eel ordered antiferromagnet.

For AFM Kitaev coupling and $\Gamma = 0.1 K$, the sharp modes at zero field are $K$- and $K_1$-modes; see Fig.~\ref{fig:AFM-Kitaev-zero-field}(c). At finite external field the resulting magnetic instability depends on field directions. Here a small $\mathbf{B}\parallel \mathbf{e}_c$ lifts the $K_1$-modes but makes the $K$-modes condense ($120^\circ$ phase), as is seen from Fig.~\ref{fig:AFM-Kitaev-field}(a). In Fig.~\ref{fig:AFM-Kitaev-field}(b), the magnetic field along the bond direction $\mathbf{e}_b$  destabilizes both modes. But the KQSL goes into an AFM phase at $\mathbf{B}=0.085K \mathbf{e}_b$ as the $K_1$-modes condense first.

\subsection{Field-induced quantum spin liquid}

Instead of weakening the KQSL the magnetic field can also have the opposite effect. Namely, in specific parameter regimes, the magnetic field strengthens the KQSL, as  shown in Fig.~\ref{fig:schematic}(c) for FM $K<0$ and $J=0.11|K|$; see Fig.~\ref{fig:proximate-KQSL} below. In our example, the modes at the $M$-points are condensed at zero field but are lifted by the perpendicular magnetic field to finite frequency. For even higher fields a much weaker mode condenses at $K$-points. This suggests that the zigzag order at zero field is destroyed as the magnetic field is switched on, and the system becomes a KQSL for intermediate field values. Therefore, we have found an explicit illustration of the `field-induced quantum spin liquid' scenario discussed in the Introduction. The possibility of such a phase diagram is shown schematically in Fig.~\ref{fig:phase-schematic}(a).

\section{Conclusions}
\label{sec:concl}

In this paper, we have developed a self-consistent RPA theory for the dynamical response of interacting Majorana fermion systems computing the generalized susceptibility  by summing over the geometric series of one-loop diagrams. We then applied the method to study the INS response of a KQSL given by the $KJ\Gamma$-model~\eqref{eq:spin-Hamiltonian} under an external magnetic field. Within a parton MFT approximation, we obtain the spectrum of Majorana fermions and use the self-consistent RPA theory for calculating the spin susceptibility and associated INS cross-section. For the pure Kitaev model without a field we benchmarked our theory with the exact calculation, which shows remarkable agreement. Comparing it to the very accurate 'adiabatic approximation' of Refs.~\cite{Knolle2014,knolle2015dynamics} we elucidate how the interactions between Majorana formions treated in RPA mimic the effect of local flux excitations, which are crucial in the exact solution. 

Beyond the pure Kitaev points we find that non-zero $J$ and $\Gamma$ interactions generically can lead to sharp para-magnon like modes coexisting with the broad Majorana continuum. These modes can be interpreted as quasi-particle poles of the collective spin excitations. Put simply, these sharp spin flip excitations are bound-states of pairs of Majoranas or 'spinon excitons'. The latter can then interact with the two-Majorana continuum and either experience level repulsion or become overdamped. 
In our calculations we also find that for increasing interactions these modes condense, allowing us to deduce the magnetic phase diagram in the vicinity of the KQSL which we find in qualitative agreement with the ED results~\cite{Rau2014}. 

Generally, a finite external magnetic field tends to destabilize the KQSL into one of the underlying magnetically ordered phases. The effect is more pronounced for the field along the bond direction $\mathbf{e}_b$, since the Majorana spectrum remains gapless in this case and tends to be more susceptible to interaction-induced instabilities. For the same reason, the KQSL under magnetic fields along the perpendicular $\mathbf{e}_c$-direction tends to be more robust due to the field-induced $c$-Majorana gap. We also find that, with certain values of $J$ and $\Gamma$, the magnetic field can stabilize the KQSL realizing the `field-induced KQSL' scenario: the sharp mode which condenses at a high-symmetry momenta at zero field is lifted to finite frequency upon increasing field.

Our theory can be readily compared with experiment but despite the qualitative success in reproducing the response at the souble Kitaev point and phase diagram found in numerics, we do not expect the MFT+RPA to give quantitatively correct results. Nevertheless, one can crisply understand a number of qualitative features. First, the existence of sharp modes in frequency does not necessarily point to an underlying simple long-range ordered state. Second, the broadening of these modes as a function of field strength can be used to understand the coupling to a continuum of fractionalized excitations. Third, strong intensity can dominate away from the high symmetry point in the BZ which calls for careful INS experiments for example for $\alpha$-RuCl$_3$, which could hopefully also help to establish a microscopic Hamiltonian description. 

Our work opens a number of paths for future research. First, in the context of Kitaev materials, the sharp INS modes appear similar to conventional  magnon excitations. However, we expect that their broadening via interaction with a continuum of fractionlaized excitations is qualitatively distinct from magnon breakdown via scattering with a two magnon continuum~\cite{winter2017breakdown}. For example, basic kinematics determine the two-magnon continuum from the single magon modes, which is not the case for the Majorana continuum. As a result, the two magnon continuum moves up in energy in a systematic way for increasing magnetic field but the Majorana continuum can remain gapless and change in a non-systematic way depending on microscopic exchange strengths.  Second, on a broader conceptual level, the surprising agreement between the exact solution of the pure Kitaev model and our theory confirms the power of the self-consistent RPA method. It can be applied to a whole range of other frustrated spin models where we have good parton MFT descriptions of the ground states, e.g.\ established via variational Monte Carlo schemes~\cite{iqbal2016spin,iqbal2011projected}. In addition, it will be interesting to investigate for U(1) spin liquids the interplay between gauge fluctuations and spinon interactions~\cite{balents2020collective}. Third, it will be very worthwhile to develop similar schemes for other dynamical response functions like Raman scattering and the dynamical polarization response. Finally, beyond the QSL context, one can employ our method to calculate the spectra of one dimensional critical spin chains in the Majorana representation~\cite{shastry1997majorana} or interacting Majorana models relevant for for topological superconductors~\cite{rahmani2015phase}. 

In conclusion, the RPA approximation has been hugely successful in describing collective excitations of interacting electron systems like metals or quantum chemistry few body systems~\cite{ren2012random}. We have shown its success in describing the dynamical response of interacting Majorana systems, which often emerge as effective descriptions of correlated quantum liquids. It opens the door for calculating dynamical response functions in regimes which are challenging for state-of-the-art numerical methods. An advantage is its simple structure allowing for an intuitive interpretations which will be key for understanding experimental response functions, for example as measured in INS.

\acknowledgments{We would like to thank K. Dixit, A. Banerjee, C. Balz, S. Nagler and A. Tennant for related experimental collaboration. We thank J. Willsher and O. Starykh for helpful discussions. P.R. is grateful to J. Habel and H.K. Jin for very useful advice on numerical methods.
J.K. also thanks the hospitality of Aspen Center for Physics, which is supported by National Science Foundation grant PHY-2210452. JK acknowledges support from the Deutsche Forschungsgemeinschaft (DFG, German Research Foundation) under Germany’s Excellence Strategy (EXC–2111–390814868 and ct.qmat EXC-2147-390858490),
and DFG Grants No. KN1254/1-2, KN1254/2-1 TRR 360 - 492547816 and SFB 1143 (project-id 247310070), as well as the Munich Quantum Valley, which is supported by the Bavarian state government with funds from the Hightech Agenda Bayern Plus. J.K. further acknowledges support from the Imperial-TUM flagship partnership.} 

Code for numerically simulating the spin susceptibility of the $KJ\Gamma$-model is available upon reasonable request~\cite{code}.

\nobalance

\appendix

\section{Majorana Mean Field theory for the $KJ\Gamma$-model}\label{sec:MajoranaMF}

\subsection{The MF Hamiltonian}

The MFT Hamiltonian consists of bilinears of $c$-Majoranas and $b$-Majoranas respectively, whereas the magnetisation and external field induce mixing between $c$ and $b$. In real space we have:
\begin{equation}
    H_{\text{MFT}} = H_c+H_b + H_m -i B^\alpha \sum_i c_i b^\alpha_i+ \sum_{i,\alpha} \lambda_\alpha F^\alpha_i\label{eq:MF-Hamiltonian-Majorana}
\end{equation}
with the $c$-Majorana Hamiltonian:
\begin{equation}
    H_c =  -i \sum_{i,j \in \alpha } \left[(K+J)Q^{\alpha \alpha}_{\alpha} + J \sum_{\beta \ne \alpha} Q^{\beta \beta}_{\alpha}+ \Gamma \sum_{\beta \ne\gamma\ne \alpha}Q^{\beta \gamma}_{\alpha}\right] c_i c_j.
\end{equation}
The $b$-Majorana part is given by:
\begin{equation}
    H_b =  - i\sum_{i,j \in \alpha } \eta_\alpha \left[(K+J) b^{\alpha}_ib^{\alpha}_j + J\sum_{\beta \ne \alpha} b^{\beta}_ib^{\beta}_j + \Gamma  \sum_{\beta \ne\gamma\ne \alpha} b^{\beta}_ib^{\gamma}_j\right].
\end{equation}
The coupling to magnetisation reads
\begin{equation}
\begin{split}
   &H_m = \sum_{i,j\in \alpha} \bigg[ (K+J)  m^{\alpha} i c_ib^\alpha_i+ J \sum_{\beta \ne \alpha}  m^{\beta}i c_ib^\beta_i \\
   &+\Gamma \sum_{\beta \ne \gamma \ne \alpha} \left(  m^\gamma ic_ib_i^\beta  + m^\beta ic_ib_i^\gamma + m^\beta i c_jb_j^\gamma + m^\gamma ic_jb_j^\beta \right)  \bigg].  
\end{split}
\end{equation}

\begin{figure*}
    \centering
    \includegraphics[width=\linewidth]{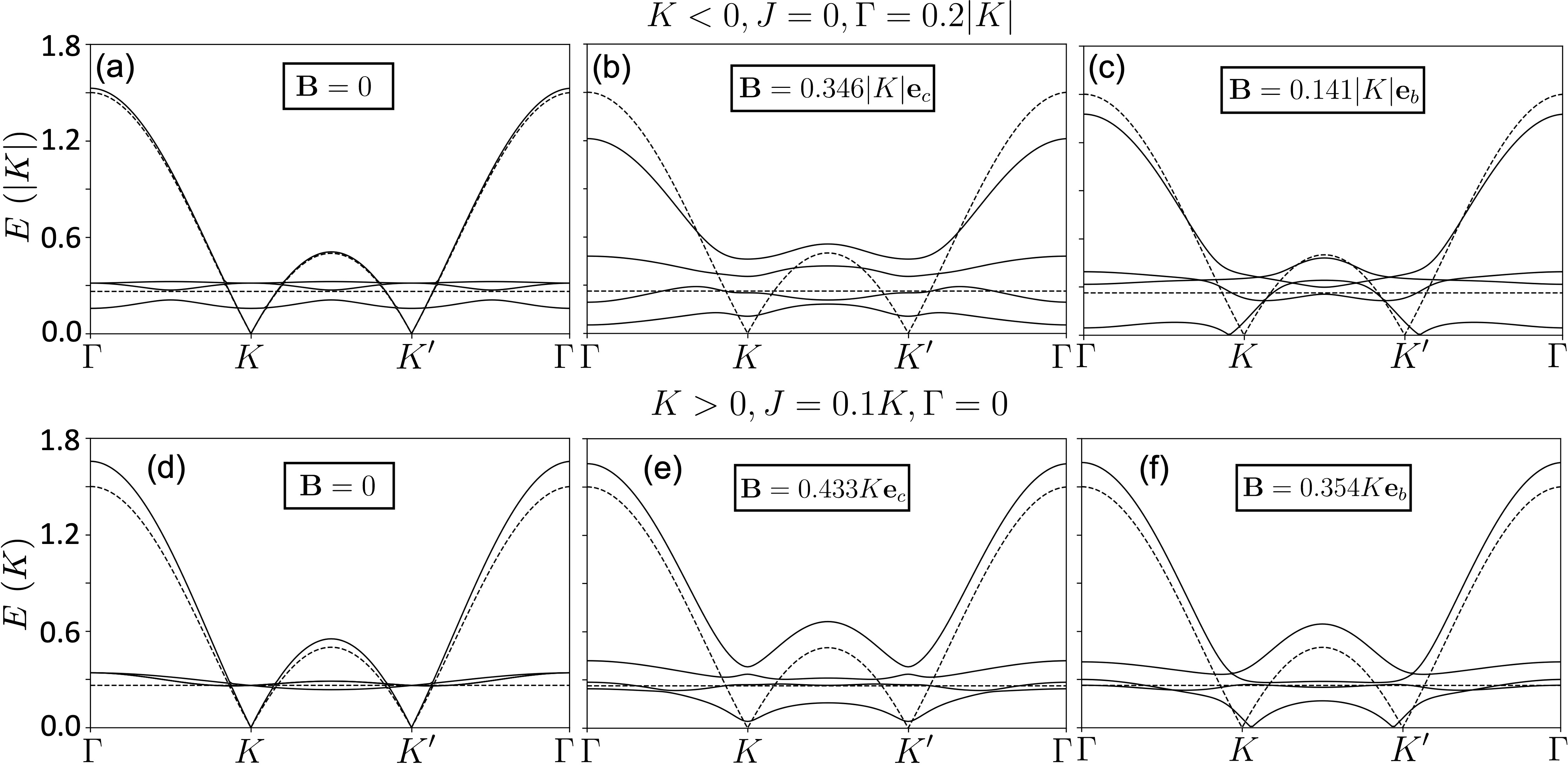}
    \caption{The MFT Majorana spectrum of the $KJ\Gamma$-model \eqref{eq:spin-Hamiltonian} at various external field values for: (a)-(c) $K<0, J=0, \Gamma = 0.2 |K|$ at the MF critical fields $\mathbf{B}_{\text{cr}}$; (d)-(f) $K>0, J=0.1 K, \Gamma =0$. The spectrum for the pure Kitaev model at zero field is shown as dashed lines. The $c$-Majorana fermions acquire a gap for $\mathbf{B} \parallel \mathbf{e}_c$, whereas for $\mathbf{B} \parallel \mathbf{e}_b$ they remain gapless. 
    }
    \label{fig:MF-spectrum}
\end{figure*}

The MFT parameters are determined by numerically solving self-consistently Eqs.~\eqref{eq:MFDecoupling-SL}, \eqref{eq:MFDecoupling-mfield} and \eqref{eq:MFDecoupling-constraints}. First, we express the averages of Majorana bilinears in terms of eigenstates of the MFT Hamiltonian \eqref{eq:MF-Hamiltonian-Majorana}, which are in turn functions of the MF parameters. For this purpose we diagonalize the MF Hamiltonian using the method outlined in Sec.~\ref{sec:RPA}.

The MF Hamiltonian Eq.~\eqref{eq:MF-Hamiltonian-Majorana} in momentum space acquires the form Eq.~\eqref{eq:free-Hamiltonian-momentum}. This allows us to compute any averages of Majorana bilinears by using Eq.~\eqref{eq:Bogoliubov-transform}:   
\begin{align}
    i \langle \gamma_a (\mathbf{r}_i)\gamma_b (\mathbf{r}_j) \rangle = i  \sum_{n>0} \sum_{\mathbf{p}} u_{an}(\mathbf{p})u_{bn}^*(\mathbf{p})e^{i\mathbf{p}.(\mathbf{r}_{i}-\mathbf{r}_j)},\label{eq:Majorana-averages}
\end{align}
where the summation is taken over the first BZ and particle bands only. The MF parameters are determined by numerically solving the combined equations Eq.~\eqref{eq:Majorana-averages} for MF averages, taking $u_{an}(\mathbf{p})$ as functions of the MF parameters from exact diagonalization. In numerical implementation, the first BZ is chosen to be the parallelogram formed by the reciprocal lattice vectors $\mathbf{b}_1, \mathbf{b}_2$ as shown in Fig.~\ref{fig:lattice}(c). A given momentum is then represented as:
\begin{equation}
\mathbf{p} =\frac{1}{L} \left(n_1 \mathbf{b}_1 + n_2 \mathbf{b}_2\right),\label{eq:momentum-grid}
\end{equation} 
where $n_1, n_2 < L$ are positive integers and $L$ determines the momentum grid interval. In all MF calculations $L=60$.

\subsection{\label{sec:MFsymmetry}Symmetry of MFT parameters}

The Majorana operators transform projectively under lattice symmetry operations. However, it is assumed that physical symmetry is not broken spontaneously by the KQSL ground state and the MF parameters retain the physical symmetry of the system~\cite{You2012}. As mentioned earlier, we consider the high-symmetry directions $\mathbf{B}\parallel\mathbf{e}_c$ and $\mathbf{B}\parallel\mathbf{e}_b$ in this paper. The symmetries then impose certain relations between the MF parameters. Below, we shall write down the independent MFT parameters. Note that from the form of constraints in Eq.~\eqref{eq:MF-Hamiltonian-Majorana} it can be seen that $\lambda_\alpha$ transform like components of a pseudovector under symmetry operations like spin, and not all of them are independent. However, in numerical calculations we keep all $\lambda_\alpha$ to verify the system symmetry.

For $\mathbf{B}\parallel\mathbf{e}_c$ there are $7$ independent MF parameters:
\begin{align}
    \eta_{x}; \ Q^{xx}_{x},  Q^{yy}_{x}, \ Q^{zz}_{x}, \ Q^{xy}_{z};\ \mathbf{m} = m\mathbf{e}_c; \lambda_x.
\end{align}
Other MF parameters are obtained by applying $C_6$ rotations successively. 

For $\mathbf{B}\parallel\mathbf{e}_b$, the number of independent MF parameters increases to $13$:
\begin{equation}
\begin{split} 
    &\eta_{x}, \ \eta_{z}; Q^{xx}_x, Q^{zz}_z,  Q^{yy}_x, Q^{zz}_x,   Q^{xx}_z Q^{xy}_z, Q^{xz}_y,  Q^{yx}_z, Q^{yz}_x; \\
    &\mathbf{m}= m\mathbf{e}_b; \lambda_x.
\end{split}    
\end{equation}
Other MF parameters are obtained by applying the in-plane reflection $\sigma$. We verify that $\lambda_z=0$ which is enforced by symmetry.

\subsection{\label{sec:MFspectrum}MFT spectrum}

In Fig.~\ref{fig:MF-spectrum}, we show the Majorana spectrum for various values of $K, J$ and $\Gamma$. We observe that away from the pure Kitaev limit at zero field, the static $b$-Majorana bands become dispersive, whereas the $c$-Majorana fermions remain gapless. Compared to $J$, the term $\Gamma$ hybridizes different $b^\alpha$-Majorana fermions and causes stronger level repulsion between the $b^\alpha$-Majorana bands. 

The effect of an external magnetic field on the spectrum is similar to the pure Kitaev model case. A small finite magnetic field along $\mathbf{e}_c$ direction induces a gap for the $c$-Majorana band. For small fields along the $\mathbf{e}_b$ direction, the $c$-Majoranas remain gapless although the Dirac cones are shifted due to the breaking of $C_6$ symmetry. This agrees with previous studies of the $KJ\Gamma$-model based on third-order perturbation theory~\cite{Takikawa2019,Hwang2022}. Also note the much more pronounced effect of the magnetic field for FM Kitaev coupling, in particular, the decrease of Majorana bandwidth as magnetic field increases. This is in agreement with the fact that the KQSL for FM Kitaev coupling is much less robust under magnetic field. As the magnetic field increases, the system undergoes a first-order MF transition into a field-polarized state, in which all spin liquid MF parameters vanish and the magnetization $|\mathbf{m}| = 1/2$ converges to the maximum value. All Majorana bands become flat indicating the breakdown of the MFT and any KQSL correlations.

\subsection{SCBA approximation for the Majorana Green's functions}\label{sec:MajoranaMF:SCBA}

As mentioned in Sec.~\ref{sec:RPA}, the MF Majorana Green's functions are obtained in the self-consistent Born approximation given by Fig.~\ref{fig:susceptibility-diagram}(b). We demonstrate this concretely in the KQSL. For example, consider the self-energy for $c$-Majorana fermions of the pure Kitaev model in real space. The Dyson's equation in Fig.~\ref{fig:susceptibility-diagram}(c) gives:
\begin{equation}
  \Sigma_{cc}(\tau;\mathbf{r}_i,\mathbf{r}_j ) = K\sum_\alpha \delta(\mathbf{r}_i-\mathbf{r}_j+\mathbf{e}_\alpha) \langle  b^\alpha(\tau,\mathbf{r}_i)  b^\alpha(\tau,\mathbf{r}_j)  \rangle,
\end{equation}
which recovers the MFT result. Other channels are obtained from the corresponding vertices, while the Lagrange multipliers for constraints enter as constant external potentials with their saddle-point values. 

\begin{widetext}
\section{Spin susceptibilities of the Kitaev QSL}\label{sec:KQSL-formulae}

In this Appendix we simplify the generalized susceptibility formulae in Eqs.~\eqref{eq:generalised-one-loop-result} and \eqref{eq:RPA-generalised-susceptibility} for the spin susceptibility of KQSL.

As discussed in Sec.~\ref{sec:RPA-spin-susceptibility}, instead of the tensor Eq.~\eqref{eq:generalised-susceptibility-one-loop} it is more convenient to define the one-loop spin-spin correlation function:
\begin{subequations} \label{eq:susceptibility-one-loop}
\begin{align}
    &[\chi_0(\omega,\mathbf{k})]_{i\alpha,j\beta}  = \lim_{i\omega_0 \to \omega +i\delta }\sum_{k,l} \int \mathrm{d}\tau P^{(0)}_{i\alpha,j\beta}(\tau,\mathbf{r}_{k} -\mathbf{r}_l) e^{i \omega_0 \tau -i\mathbf{k}.(\mathbf{r}_{k} -\mathbf{r}_l)}; \\
    &P^{(0)}_{i\alpha,j\beta}(\tau,\mathbf{r}) = \langle \text{T}_\tau \{i c_i(\tau,\mathbf{r}) b^\alpha_i(\tau,\mathbf{r}) i c_j(0,0) b^\beta_j(0,0) \}\rangle_0.
\end{align}    
\end{subequations}
where $i,j$ are sublattice indices. Averaging of Eq.~\eqref{eq:susceptibility-one-loop} over the MFT ground state gives:
\begin{equation}\label{eq:one-loop-1}
\begin{split}
    [\chi_0(k)]_{i\alpha,j\beta}=&-\sum_{\mathbf{p}} \int \mathrm{d}\tau e^{ik_0 \tau} \Big[ \bigl\langle \text{T}_\tau \{c_i(\tau ,\mathbf{p}+\mathbf{k}) \overline{c}_j(0,\mathbf{p}+\mathbf{k})\} \bigl \rangle \bigl\langle\text{T}_\tau \{ b_j^\beta(0 ,\mathbf{p}) \overline{b}_i^\alpha(\tau ,\mathbf{p})\}\bigl\rangle \\
    &-\bigl\langle \text{T}_\tau \{ c_i(\tau ,\mathbf{p}+\mathbf{k})\overline{b}_j^\beta(0 ,\mathbf{p}+\mathbf{k})  \} \bigl\rangle \bigl\langle \text{T}_\tau \{ c_j(0 ,\mathbf{p})  \overline{b}^\alpha_i(\tau ,\mathbf{p}) \}\bigl\rangle \Big].    
\end{split}   
\end{equation}
Again summing over internal Matsubara frequency and performing the analytical continuation, Eq.~\eqref{eq:generalised-one-loop-result} now becomes explicitly:
\begin{equation}
\begin{split}\label{eq:one-loop-result}
    &[\chi_0(k)]_{i\alpha,j\beta}   
    =-\sum_{\mathbf{p}} \sum_{m,n} \Biggl[ \frac{u^c_{in}(\mathbf{p}+\mathbf{k})u^{c*}_{jn}(\mathbf{p}+\mathbf{k})u^{\beta*}_{jm}(-\mathbf{p})u^\alpha_{im}(-\mathbf{p})}{\omega -E_n(\mathbf{p}+\mathbf{k})-E_m(-\mathbf{p})+  i\delta}  
    -  \frac{u^{c*}_{in}(-\mathbf{p}-\mathbf{k}) u^c_{jn}(-\mathbf{p}-\mathbf{k})u^{\beta}_{jm}(\mathbf{p})u^{\alpha *}_{im}(\mathbf{p})}{\omega +E_n(-\mathbf{p}-\mathbf{k})+E_m(\mathbf{p})+  i\delta} \\
    &  - \frac{u^c_{in}(\mathbf{p}+\mathbf{k})u^{\beta *}_{jn}(\mathbf{p}+\mathbf{k})u^{c*}_{jm}(-\mathbf{p})u^{\alpha}_{im}(-\mathbf{p})}{\omega -E_n(\mathbf{p}+\mathbf{k})-E_m(-\mathbf{p})+  i\delta} 
    + \frac{u^{c*}_{in}(-\mathbf{p}-\mathbf{k}) u^{\beta}_{jn}(-\mathbf{p}-\mathbf{k})u^c_{jm}(\mathbf{p})u^{\alpha *}_{im}(\mathbf{p})}{\omega +E_n(-\mathbf{p}-\mathbf{k})+ E_m(\mathbf{p})+ i\delta} \Biggl].
\end{split}
\end{equation}
\end{widetext}
We have written $u^c_{in}$ to represent the Bogoliubov coefficients $u_{an}$ in Eq.~\eqref{eq:Bogoliubov-transform} where the internal index $a$ includes $c$-Majoranas on sublattice $i=A, B$; $u^\alpha_{in}$ is written similarly for $b^\alpha$-Majoranas.

The RPA susceptibility is now given by:
\begin{equation}
     [\chi(\omega,\mathbf{k})]_{i\alpha,j\beta} = [\chi_0(\omega,\mathbf{k})]_{i\alpha,k\gamma}[1 + \hat{U} (\mathbf{k})\chi_0(\omega,\mathbf{k})]^{-1}_{k\gamma,j\beta},
     \label{eq:RPA-tensor}
\end{equation}
in terms of the symmetrized interaction vertex:
\begin{equation}
    [\hat{U}(\mathbf{k})]_{i\alpha,j\beta}  = U_{i\alpha,j\beta} (-\mathbf{k}) + U_{j\beta,i\alpha}(\mathbf{k}),
\end{equation}
where $U_{i\alpha,j\beta}  (\mathbf{k})$ is defined in Eq.~\eqref{eq:interaction-vertex} in the main text.

%

\end{document}